\documentclass[12pt,a4paper]{article}

\usepackage[utf8]{inputenc}
\usepackage[T1]{fontenc}
\usepackage{amsmath,amssymb,mathrsfs,bm}
\usepackage{graphicx}
\usepackage[numbers,sort&compress]{natbib}
\usepackage{xcolor}
\usepackage[margin=2.5cm]{geometry}
\usepackage{setspace}
\usepackage{dcolumn}
\usepackage{times}
\usepackage{caption}
\usepackage{hyperref}
\usepackage[yyyymmdd]{datetime}

\usepackage{docmute}   

\let\oldaddcontentsline\addcontentsline

\begin{document}

\renewcommand{\addcontentsline}[3]{}

\title{\Large\bfseries Quantum-memory-assisted on-demand microwave-optical transduction}

\author{
Hai-Tao Tu$^{1,2,3,\dagger}$, 
Kai-Yu Liao$^{1,2,3,\dagger,\ast}$, 
Si-Yuan Qiu$^{1,2,\dagger}$, 
Xiao-Hong Liu$^{1,2}$, \\
Yi-Qi Guo$^{1,2}$, 
Zheng-Qi Du$^{1,2}$, 
Yang Xu$^{1,2}$, 
Xin-Ding Zhang$^{1,2}$, \\
Hui Yan$^{1,2,4,5,\ast}$, 
Shi-Liang Zhu$^{1,2,3,5,\ast}$
}
\date{}
\maketitle

\noindent
$^{1}$Key Laboratory of Atomic and Subatomic Structure and Quantum Control (Ministry of Education), Guangdong Basic Research Center of Excellence for Structure and Fundamental Interactions of Matter, and School of Physics, South China Normal University, Guangzhou 510006, China

\noindent
$^{2}$Guangdong Provincial Key Laboratory of Quantum Engineering and Quantum Materials, Guangdong-Hong Kong Joint Laboratory of Quantum Matter, and Frontier Research Institute for Physics, South China Normal University, Guangzhou 510006, China

\noindent
$^{3}$Quantum Science Center of Guangdong-Hong Kong-Macao Greater Bay Area, Shenzhen, China

\noindent
$^{4}$GPETR Center for Quantum Precision Measurement, South China Normal University, Guangzhou 510006, China

\noindent
$^{5}$Hefei National Laboratory, Hefei 230088, China

\bigskip
\noindent
$^{\dagger}$These authors contributed equally to this work.

\noindent
$^{\ast}$Email: kaiyu.liao@m.scnu.edu.cn; yanhui@scnu.edu.cn; slzhu@nju.edu.cn

\begin{center}
{\large\bfseries Abstract}
\end{center}
\noindent
Microwave-optical transducers and quantum memories are essential for quantum repeaters enabling a quantum internet. Despite advances in both technologies, integrating these functionalities remains challenging. Here, we theoretically propose and experimentally demonstrate an on-demand microwave-optical quantum transducer based on a Rydberg ensemble. Using cascaded electromagnetically induced transparency, we store microwave photons in a highly excited collective state and convert them into optical photons during retrieval. Leveraging an optical depth of millions for microwave photons and minimal single-photon-level dephasing, our transducer achieves around 90\% area-normalized storage efficiency, 2.3 MHz bandwidth, and noise-equivalent temperature of 26 K under cavity-free conditions. Furthermore, our system is cryogenically compatible and extendable for high single-photon conversion efficiency without requiring optical cavity coupling. These findings advance practical on-demand quantum interfaces with broad applications across atomic and solid-state platforms.

\section{Introduction}

One of the key goals of quantum science and technology is to build a quantum internet where quantum computers serve as nodes, connected by optical fibers as data transmission lines~\cite{KimbleNature2008, WehnerSci2018}. Solid-state qubits, including those based on superconducting circuits~\cite{ClarkNature2008, DevoretScience2013}, nitrogen-vacancy centers~\cite{PompiliScience2021}, and semiconductor quantum dots~\cite{ArquerScience2021}, are promising candidates for quantum computers. If the nodes use solid-state qubits, establishing such a quantum internet requires two other essential components in addition to quantum computers and optical fibers: microwave-optical (MO) transducers and quantum memories. The solid-state quantum computers typically operate within the microwave (MW) frequency range; however, the loss of MW signals during transmission is significant. This necessitates the use of MO transducers to convert quantum information carriers, represented by MW photons, into optical photons~\cite{LambertReview2019, LaukReview2020, XiangRMP2013,XHan2021}. Since optical photons experience less loss and thermal noise, long-distance quantum information transmission can be achieved through optical fibers. The establishment of long-distance entanglement between nodes relies on the Bell-state measurement of two optical photons arriving simultaneously at the beam splitter of a middle station~\cite{Duan2001, Sangouard2011}. Synchronicity is crucial for the successful realization of entanglement in the protocol, and it is generally ensured by directing the photons into a quantum memory and retrieving them at the required time to guarantee simultaneity~\cite{LvovskyNP2009}.

Extensive research has been conducted on the aforementioned transducers and quantum memories, yielding a series of significant advances. Quantum transducers have been experimentally realized or theoretically proposed in a variety of systems, ranging from optomechanical systems~\cite{BochmannNP2013, HigginbothamNP2018, ForschNP2019}, atomic ensembles~\cite{Han2018, Tu2022, Kumar2023, Borowka2024,Vogt2019}, electro-optical systems~\cite{RuedaOptica2016, FanSA2018}, to magnonic systems~\cite{WilliamsonPRL2014}. Typically, quantum transduction requires high efficiency, high fidelity, and broad bandwidth. By utilizing an intermediate system that coherently couples to both MW and optical modes in steady states, all of these demonstrations must adhere to direct conversion (DC) protocols, where heralded MW photons are instantly converted into optical signals without delays. However, DC schemes face issues such as atomic heating and additional optical noise introduced by continuous-wave optical pumping during the conversion process~\cite{Kurokawa2022}. On the other hand, quantum memory is regarded as a vital element in quantum information processing, serving as a synchronization tool that aligns various processes in quantum computing, a device for converting heralded photons into on-demand photons, and a buffer for temporally isolating intensive optical pumping. Quantum memories have also been experimentally implemented in atomic ensembles and doped rare-earth solid-state systems~\cite{Saito2013, Julsgaard2013, Ranjan2020, Flurin2015, Reagor2016, Pfaff2017, Bao2021, Palomaki2013, Liu2023, Hsiao2018, Vernaz2018, Distante2017, Wang2019,KYSu2022}. Currently, there is a pressing need to develop quantum memories that can achieve both high efficiency and long-duration storage. Of course, integrating these components to perform various functions remains a highly challenging task. The development of a single device capable of multiple functions is an important direction for advancement. However, the functional integration of an absorptive quantum memory with a MO transducer, which is crucial for suppressing pump-induced noise and enabling on-demand operation, remains experimentally unrealized.

In this article, we theoretically propose and experimentally demonstrate a quantum-memory-assisted on-demand microwave-optical quantum transducer (OMQT), which is intrinsically suitable for efficient quantum networking between the remote solid-state qubits. In this context, `on-demand' means that the transduction process produces photons at a precisely controlled time, rather than at random or probabilistic times. This capability is enabled by the quantum memory, which converts heralded MW photons into photons that are released exactly when needed for subsequent quantum operations. By employing the multichannel electromagnetically induced transparency (EIT) process, this scheme integrates conversion and quantum memory capabilities based on a Rydberg ensemble. We demonstrate an experimental implementation of our scheme in the laser-cooled atoms, by storing MW photons in a highly excited collective state and subsequently converting them into optical photons during the retrieval. Benefiting from the optical depth (OD) with order of millions for MW photons in Rydberg ensemble, we establish optimal control strategies for the storage and retrieval of photon wave packets. We present a proof-of-concept transducer with an area-normalized storage efficiency (ASE), defined as the ratio of recalled optical photons to input MW photons, up to 90\% at a storage time of 50 ns, and still above the 50\% quantum no-cloning threshold at 0.56 $\mu$s, along with a conversion bandwidth of 2.3 MHz and a noise-equivalent temperature as low as 26 K under cavity-free conditions. We also observe stored-photon-number dependent Rydberg dephasing. Our work paves the way to an efficient quantum interface between atomic and solid-state qubits.

\section{Results}

\subsection{OMQT scheme}

We first present our OMQT scheme, which can be applied to a quantum network linking remote superconducting qubits. In this scheme, a cigar-shaped ensemble of cold atoms is trapped at a distance of 10--20~$\mu$m from the surface of a coplanar waveguide (CPW)~\cite{XiangRMP2013, KiffnerNJP2016, FortaghPRR2022}, which is connected to superconducting qubits. Assuming the CPW has a strip width of $w \simeq 30$~$\mu$m, an electrode distance of $d \simeq 25$~$\mu$m, and an effective dielectric constant $\epsilon_{r}\simeq 6$, higher-order modes of the CPW can be effectively suppressed~\cite{FortaghNC2017, FortaghPRapp2025}, and MW photons propagate in TEM mode along the waveguide. To characterize the imperfect spatial overlap between the atomic ensemble and the TEM mode, we define a filling factor as $F = A_\mathrm{L}/A_\mathrm{w}$, where $A_\mathrm{L}$ ($A_\mathrm{w}$) is the transverse size of the auxiliary laser beam (waveguide mode).

\begin{figure}[p]
\begin{center}
\includegraphics[width=15.0cm]{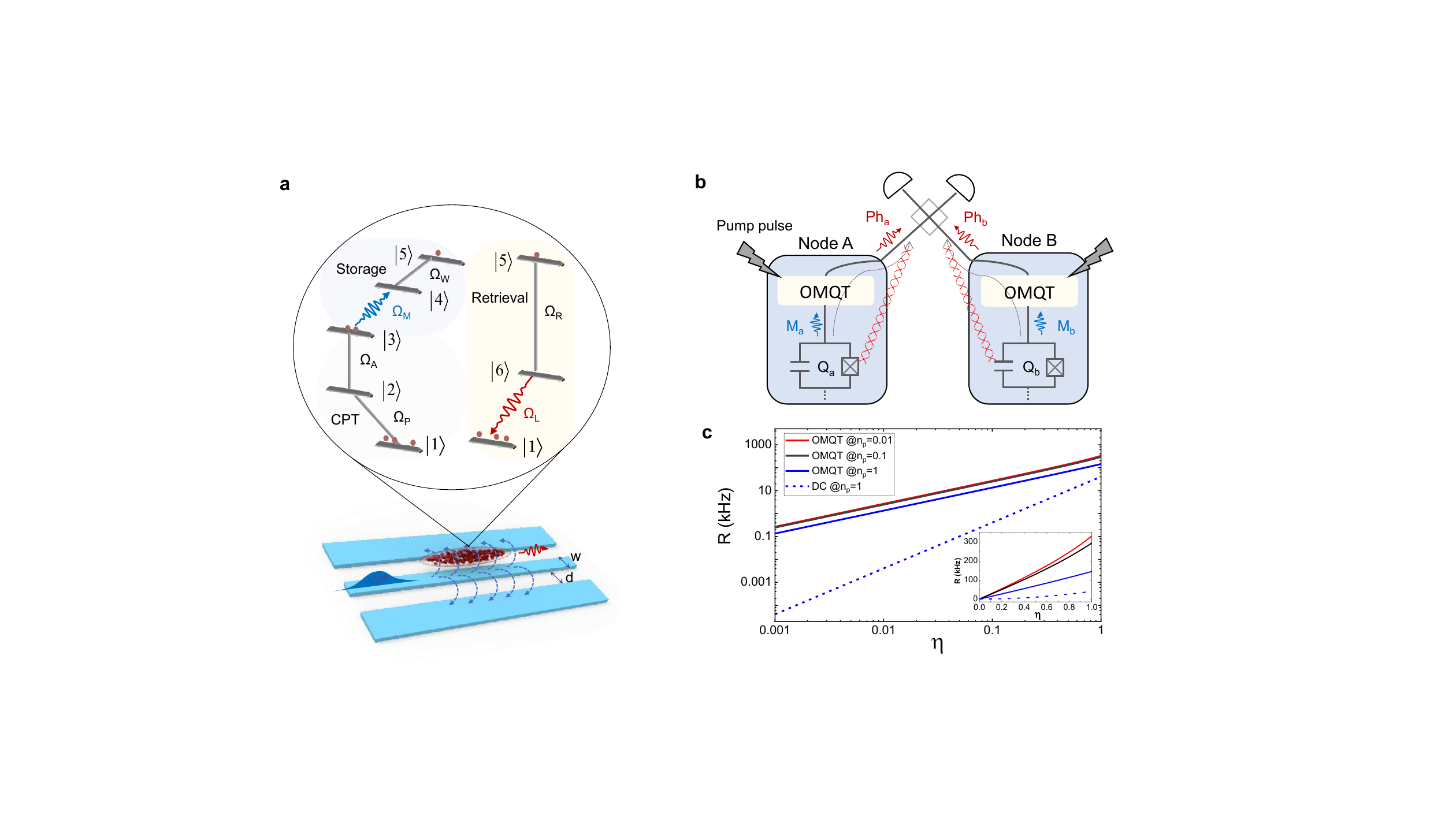}
\caption{\label{fig:1} \textbf{OMQT scheme and its application for entanglement generation.} \textbf{a}, A cigar-shaped atomic ensemble trapped above the CPW serves as an interface for implementing the OMQT scheme. The dashed line represents electric field distribution of TEM mode. The inset shows the energy-level configuration, consisting of MW photon storage (left) and optical photon retrieval (right). \textbf{b}, Illustration of the entanglement generation between solid-state qubits Q$_{\rm{a}}$ in node $A$ and qubits Q$_{\rm{b}}$ in node $B$ using OMQTs. \textbf{c}, Entanglement generation rate $R$ between the remote solid-state qubits as a function of conversion efficiency $\eta$. The solid lines correspond to the OMQT scheme for different noise photon numbers $n_\mathrm{p}$, and the dashed line is that of the direct conversion scheme. The inset illustrates the same plot in a linear scale.}
\end{center}
\end{figure}

The conversion mechanism can be regarded as the process of multi-level quantum memory consisting of two sets of EIT mediated by a Rydberg ensemble. The schematic of OMQT and relevant atomic levels are illustrated in Fig.~\ref{fig:1}a. During the storage stage, atomic population is initially coherently trapped in a superposition state encompassing the ground level $|1\rangle$ and the Rydberg level $|3\rangle$. With the aid of dressed states, MW photons that resonate with the Rydberg transition $|3\rangle \rightarrow |4\rangle$ are coupled into the ensemble under cascade-type EIT conditions while the write field is activated. Subsequently, the MW photon is absorbed and stored as a collective excitation in Rydberg state $|5\rangle$ after the write field is turned off. Once the designated storage period concludes, the excitation is retrieved in reverse from the $|6\rangle \rightarrow |1\rangle$ transition and converted into an optical photon by applying the read field to the ensemble~\cite{Wang2019}.

Since MW photons are transversely confined, we build a one-dimensional model to theoretically predict the on-demand transduction~\cite{KiffnerNJP2016}. Under conditions of weak excitation, the transduction dynamics for the MW and optical field envelopes $\Omega_{\rm{M}}(z, t)$ and $\Omega_{\rm{L}}(z, t)$ are described by the first-order Maxwell-Bloch equations:
\begin{equation}
\label{Maxwell-Bloch}
\left(\partial _ { z } + \frac{1}{c} \partial _ { t } \right)\binom{\Omega_{\rm{M}}(z, t)}{\Omega_{\rm{L}}(z, t)}=-\frac{i}{2}  \binom{F n_3 \sigma_{\rm{M}} \Gamma_{4} P_{43}(z, t)}{n_1 \sigma_{\rm{L}} \Gamma_{6} P_{61}(z, t)},
\end{equation}
where $P_{ij}$ denotes the first-order atomic coherence associated with the transition $|i\rangle \leftrightarrow |j\rangle$, $\Gamma_{j}$ represents the total decay rate of the atomic coherence at level $|j\rangle$, $n_j$ is the number density of atoms in state $|j\rangle$, and $\sigma_{M}$ ($\sigma_{L}$) is the atomic absorption cross section for the MW (optical) transition. The parameter $d_{M} = n_3 \sigma_M L$ ($d_{L} = n_1 \sigma_L L$) denotes the OD of the MW (optical) transition for an ensemble of length $L$. The large electric dipole moment $\mu$ and long lifetime (i.e., small decay rate $\Gamma$) of Rydberg states result in a huge absorption cross section $\sigma_M = \frac{4\pi  |\mu|^2 }{\lambda_M\varepsilon_0\hbar\Gamma}$, where $\lambda_M$ is the MW wavelength and $\varepsilon_0$ is the vacuum permittivity. This yields an MW OD on the order of a million. In contrast, the OD observed for an optical photon is typically several hundred.

It is noteworthy that the maximum efficiency of EIT-based memory is solely dependent on the effective OD of medium when optimal control strategies are employed for the storage and retrieval of wave packets~\cite{Gorshkov2007a}. In the waveguide integration scheme, the effective OD refers to the optical thickness coupled to the TEM mode and characterizes the medium absorption of MW photons from the CPW. Although the medium size is transversely smaller than that of TEM mode ($A_\mathrm{L} < A_\mathrm{w}$), the effective OD (i.e.\ $Fd_{M}$) readily reach tens of thousands. This raises the question of whether optimal storage can be extended to our EIT-memory-based transduction method. If so, we could achieve high efficiency for MO transducer, benefiting from the optimal control strategies utilized in optical quantum memory~\cite{Wang2019,Hsiao2018}. To investigate this possibility, we solve Eq.~(\ref{Maxwell-Bloch}) to calculate the efficiency (see Supplementary Note~1). In the context of EIT-memory-based transduction, conversion efficiency $\eta$ is influenced by three key factors: slow-light transmission $\eta_t$, spin-wave survival efficiency $\eta_s$, and the fraction $\eta_c$ of Gaussian input pulse that is compressed into the cigar-shaped ensemble during the writing process. Thus, the conversion efficiency is expressed as $\eta = \eta_t \eta_s \eta_c$. For short storage time and large OD, both $\eta_s$ and $\eta_c$ approach unity, indicating OMQT efficiency is primarily determined by $\eta_t$. By solving Eq.~(\ref{Maxwell-Bloch}) under the conditions of forward propagation and backward retrieval of a Gaussian wave packet, we derive $\eta$ for short storage times as
\begin{equation}
\label{Efficiency}
\eta \simeq \eta_t = \eta_0 e^{-2 \gamma_{51} t_d},
\end{equation}
where $\eta_0=1/\sqrt{(1+\alpha_{M}/{F d_{M}})(1+\alpha_{L}/d_L)}$, $\gamma_{51}$ denotes the decay rate of Rydberg coherence between the $|1\rangle$ and $|5\rangle$ states (see Supplementary Note~5), $t_d$ represents the total delay time, and $\alpha_{M}$ ($\alpha_{L}$) is a dimensionless coefficient linked to MW (light) delay time, decoherence rates, and pulse full width at half maximum (FWHM) duration. Therefore, a near-unity conversion efficiency ($\eta \simeq 0.9$) is possible for the waveguide-based OMQT with the typical parameters: $F \simeq$ 0.01 ($A_\mathrm{L}\simeq$ 300 ${\rm{\mu m}}^2$, $A_\mathrm{w} \simeq$ 3 $\times 10^{4} {\rm{\mu m}}^2$), $\gamma_{51}/2\pi = 10$ kHz, $t_d = 0.5$ $\mu$s, $d_M = 1\times 10^6$, $d_L = 100$, $\alpha_M = 50$, and $\alpha_L = 0.5$.

Figure~\ref{fig:1}b shows the OMQT applications in the entanglement generation between solid-state qubit Q$_{\rm{a}}$ in node $A$ and qubit Q$_{\rm{b}}$ in node $B$. Each node comprises solid-state qubits and its respective OMQT. Initially, a Bell state is established between the solid-state qubits Q$_{\rm{x}}$ (where x = a, b) and the corresponding MW photon M$_{\rm{x}}$. The MW photon M$_{\rm{x}}$ is then directed into the OMQT$_{\rm{x}}$ for on-demand storage. Subsequently, M$_{\rm{x}}$ is up-converted into an optical photon Ph$_{\rm{x}}$, which becomes entangled with qubit Q$_{\rm{x}}$. The optical photon Ph$_{\rm{x}}$ is subsequently transmitted to a central station. At this station, a Bell-state measurement involving the photons Ph$_{\rm{a}}$ and Ph$_{\rm{b}}$ serves as a herald for the successful generation of entanglement between qubits Q$_{\rm{a}}$ and Q$_{\rm{b}}$. The on-demand capability of the OMQT is crucial, as it ensures that both optical photons reach the central station simultaneously to efficiently generate entanglement~\cite{Kurokawa2022}. To compare the direct conversion, we calculate the entanglement generation rates $R$ of OMQT and DC schemes, respectively (see Methods). As shown in Fig.~\ref{fig:1}c, the $R$ of OMQT scheme is 65 kHz at an efficiency of $\eta$ = 0.5, which is nearly ten times that of direct conversion. At the low efficiencies, the advantages owing to memory enhancement become more pronounced. Moreover, OMQT can effectively mitigate the noise photons, thereby further improving the quantum interconnect performance.

\subsection{Experimental setup}

The full scheme in Fig.~\ref{fig:1}b is extremely challenging to realize. As a first step towards this goal, we present a proof-of-principle demonstration of OMQT. Specifically, we implement a free-space OMQT using a cold ensemble of $^{87}$Rb atoms in a two-dimensional magneto-optical trap (MOT). Using a unity filling factor in Eq.~(1), we evaluate the free-space OMQT and obtain an ASE $\eta_\mathrm{I}$, i.e.\ the ratio of recalled optical photons to incident MW photons on the medium, to characterize the internal photon-to-photon conversion (see Supplementary Note~7). Fig.~\ref{fig:2} presents a simplified diagram of the experimental setup (see Supplementary Note~2), and the detailed parameters of six fields are listed in Table~\ref{tab:fields}. We conduct the experiment according to the outlined scheme. Following the loading of the MOT, the atoms are optically pumped into the ground state $|1\rangle = |5S_{1/2}, F=2, m_F=2\rangle$, and are controlled in the time sequence with a 100 Hz repetition rate as previously described~\cite{Tu2022}, as shown in Fig.~\ref{fig:2}a. During the OMQT stage, we switch on a 6.4 G bias magnetic field aligned along the $z$-direction to define the quantization axis. The maximum OD of optical transition $|1\rangle \leftrightarrow |2\rangle = |5P_{3/2}, F=3, m_F=3\rangle$ is measured to be $140 \pm 4$, from which we derive a MW OD of $7.5\times 10^5$ based on the coherent population trapping (CPT) effect (see Supplementary Note~4). Two auxiliary beams, $\Omega_A$ and $\Omega_R$, sourced from different 480 nm lasers, propagate collinearly along the $z$ axis and counterpropagate with additional auxiliary fields $\Omega_P$ and $\Omega_W$. For demonstration, the signal MW is launched into free space via a helical antenna and directed toward the atomic cloud at a $3^{\circ}$ angle relative to the $z$-axis.

By adjusting the power of the probe laser, we can significantly manipulate the population in the Rydberg state $|3\rangle = |47S_{1/2}, m_{J}=1/2\rangle$ in the presence of the auxiliary fields, thereby regulating the OD of the MW transition $|3\rangle \leftrightarrow |4\rangle = |47P_{3/2}, m_{J}=3/2\rangle$. The other relevant energy levels include $|5\rangle = |46D_{5/2}, m_{J}=1/2\rangle$ and $|6\rangle = |5P_{3/2}, F=2, m_F=1\rangle$. The experiment is conducted in an ambient environment at approximately $300$ K, where only the paraxial background radiation propagating within a tiny solid angle can be efficiently stored in our cigar-shaped atomic gas and converted into optical photons(see Supplementary Note~6). Following the backward recovery process, the optical photons $\Omega_L$ propagate along the probe laser beam, after which they are collected using a single-mode fiber and a series of optical filters. Signal photons are detected using single-photon counting modules (Perkin Elmer, SPCM-AQRH-16), and coincidence counts are recorded by a time-to-digital converter (Fast Comtec P7888) with a time bin of $1$ ns. The probability that an optical photon generated from atomic ensemble can be converted into photon counting is 39\%. A $50/50$ fiber beamsplitter is employed to conduct the Hanbury Brown-Twiss (HBT) experiment. To perform the OD measurement, the probe beam is detected by an avalanche photodiode (APD).

\subsection{Demonstration of OMQT in ambient cold atoms}

We first search the phase-matching direction through realizing a direct MO transduction via a six-wave mixing process, which is driven by the constant and all-resonant auxiliary fields, and then achieve the OMQT to benchmark the noises. The Gaussian input pulse has a FWHM of $T_p$ = 300 ns. Without the temporal isolation of the read and write process, the probe laser beam increases the stray photon noise compared with OMQT, and hence the input pulse is set with a higher mean photon number per input MW pulse $\bar N$ = 130 for a sufficient signal-to-noise ratio (SNR). The direct transduction has a relatively low efficiency ($\eta_\mathrm{I} \approx$ 32\%) mainly because it saturates when the atomic population is gradually trapped in a dark state~\cite{Han2018}, defined as $|D\rangle \propto (\Omega_\mathrm{W}^{*} \Omega_\mathrm{A}^{*}|1\rangle-\Omega_\mathrm{W}^{*} \Omega_\mathrm{P}|3\rangle+\Omega_\mathrm{M}^{*} \Omega_\mathrm{P}|5\rangle)$. Then we carry out the on-demand transducer following the time sequences described in Fig.~\ref{fig:2}a. Representative signal counts for $\bar N$ = 0.1 are illustrated in Fig.~\ref{fig:3}a (middle) for the storage periods of 50 ns and 550 ns. The total noise count after a 50 ns storage time is 0.119 per pulse, where the noise sources contain thermal photon noise ($\bar n_{th}$ $\sim$ 0.109 per pulse), Rydberg fluorescence, stray photon noise ($\bar n_{st}$ $\sim$ 0.01 per pulse), and SPCM dark counts. Therefore, among them the 92\% counts are derived from the stored thermal MW photons (see Fig.~\ref{fig:3}b and Supplementary Note~6). In the absence of input photons, the ratio of thermal counts to other intrinsic noise is 10.9, and this proportion gives rise to a noise-equivalent temperature of $T_{NE}$ = 26 K. Although the converted photons are only 267 MHz away in frequency from the pump laser beam, the temporal isolation in the storage process strongly suppresses the pump light noise. Without an additional cavity-assisted filter, our system achieves a low noise-equivalent temperature that is only half of that reported in the direct transduction scheme~\cite{Borowka2024}, where the frequency difference between the generated photons and the pump light exceeds 2 THz. The observed thermal photon noise $\bar n_{th}$ is consistent with the theoretical prediction of 0.1 photons per pulse. At cryogenic temperature below 4K, it drops to below $10^{-3}$ photons per pulse, becoming negligible(see Supplementary Note~6). We note that the small MW-reception solid angle reduces the optical photons transduced from thermal noise, it simultaneously introduces significant mode-matching losses for free-space signal photons.

Unlike the input waveform, the retrieved pulse shapes exhibit a slowly falling tail at the rear probably due to the transverse inhomogeneity of recovery process. With the shutdown of uniform write field, the input MW is homogeneously mapped in the cross section of medium under the broadband EIT condition. After turning on the focused read beam, the spin wave is recovered but the rim light backward propagates with a lower velocity than the centered light, leading to an asymmetric waveform of retrieved signal. The same phenomenon is also observed for the recovered waveform of the thermal MW photons, as shown in Fig.~\ref{fig:3}a (bottom).

\subsection{Storage time}

Fig.~\ref{fig:3}c shows the measured $\eta_\mathrm{I}$ as a function of the storage time (see Methods), which is defined as the time difference between turning the control field off and then on~\cite{RMQM1, RMQM2}. It shows the storage efficiency is up to 90\% at a storage time of 50 ns. The data were fitted with a Gaussian decay function, $\exp(-\tau^2/\tau^{2}_{\mathrm{coh}})$, yielding a coherence time of $\tau_{\mathrm{coh}} \simeq 0.9~\mu\mathrm{s}$. This coherence time is slightly shorter than the lifetime of Rydberg spin-wave~\cite{Li2016,Distante2017}, likely due to atom-atom interactions and stray electric fields. To account for the exponential decay effect, we also fitted a curve form $A\exp(-\tau/\tau_{D})$ with $A = 0.93$ and $\tau_{D} = 2~\mu\mathrm{s}$. For the Gaussian (exponential) fit, the fitted $\eta_\mathrm{I}(\tau=0)$ was $82\%$ ($93\%$), and the storage time at the threshold efficiency of the quantum no-cloning regime ($\eta_\mathrm{I} = 50\%$) was $0.58$ ($0.56$) $\mu\mathrm{s}$. If we turn off the input MW, the noise counts retrieved from surrounding thermal photons display a similar decay trend, as shown in the inset of Fig.~\ref{fig:3}c, corresponding to a coherence time of 1.1 (0.9) $\mu$s.

\subsection{Conversion band and autocorrelation measurement}

Next we scan the MW detuning $\delta_M$ across the $|3\rangle\leftrightarrow|4\rangle$ transition and measure the transduction's dependence on the MW frequency at the single-photon level [$\bar N = 0.3$, see Fig.~\ref{fig:4}a]. This reveals a FWHM bandwidth of 2.3 MHz and a fitted efficiency $\eta_\mathrm{I}(\delta_M=0) = 88\%$, obtained by fitting a Gaussian function with an offset of $0.06(3)$. This bandwidth represents a twofold improvement compared to the previous direct transduction in cold atoms~\cite{Tu2022}. The offset is mainly due to the subtracted noise counts, which are approximately 0.12, as derived from Fig.~\ref{fig:3}b. To confirm that the thermal background is currently the primary noise source, we measure the second-order autocorrelation function of the converted photons $g^{(2)}(\tau)$ with a HBT experiment, and the results are plotted in Fig.~\ref{fig:4}b. In the absence of coherent MW, the measured $g^{(2)}(0) \approx$ 1.8 reflects on the bunching effect of thermal photons. The parametric fitting of $g^{(2)}(\tau)$ with an exponential function: $1 + (g^{(2)}(0)- 1)\exp(-\tau/\tau_{coh})$, determines a coherence time $\tau_{coh}$ = 0.82 $\mu$s, which is consistent with the storage time of thermal photons as shown in the inset of Fig.~\ref{fig:3}c. Then we change the strength of injected coherent MW pulse. $g^{(2)}(0)$ in Fig.~\ref{fig:4}b for $\bar{N}=5,20\ \bar{n}_{\mathrm{th}}/\eta_\mathrm{I}$ falls to 1 as the input coherent photon number outstrips the thermal photons~\cite{Kumar2023}. The measured $g^{(2)}(\tau)$ has excellent agreement with the parameter-free theoretical prediction:
\begin{equation}
g^{(2)}(\tau)=1+\frac{\left|\bar{n}_{\mathrm{th}} g_{\mathrm{th}}^{(1)}(\tau)+\eta_\mathrm{I}\bar{N} \right|^2-\eta_\mathrm{I}^{2}\bar{N}^2}{\left(\bar{n}_{\mathrm{th}}+\bar{n}_{\text {st}}+\eta_\mathrm{I}\bar{N}\right)^2}.
\end{equation}
Here the first order autocorrelation function $g_{\mathrm{th}}^{(1)}(\tau)=\frac{1}{2 \pi} \int_{-\infty}^{\infty}|S(\delta_{\mathrm{M}})|^2 \mathrm{e}^{-i \delta_{\mathrm{M}} \tau} \mathrm{d} \delta_{\mathrm{M}}$ is calculated via the Wiener--Khinchin theorem. $|S(\delta_{\mathrm{M}})|^2$ represents the normalized power spectral density, which has the same form shown in Fig.~\ref{fig:4}a. The autocorrelation result suggests that the OMQT maintains good coherence when applied to the single-photon MW pulse even under ambient conditions.

\subsection{Observation of Rydberg dephasing associated with stored photon number}

According to Eq.~(\ref{Efficiency}), the Rydberg decoherence $\gamma_{51}$ is a dominant term of degrading the storage efficiency for short storage times and high OD. Since spontaneous Rydberg decay and other dephasing mechanisms (such as finite laser linewidth and blackbody radiation) are insignificant, the primary source of Rydberg decoherence comes from dispersion in the relevant Rydberg energy shifts. We obtain that the Rydberg decoherence $\gamma_{51} = \sqrt{\bar{N}}\gamma_0$ under the mean field approximation(see Supplementary Note~5), and then Eq.~(\ref{Efficiency}) is rewritten as
\begin{equation}
\label{Efficiency2}
\eta_\mathrm{I} \simeq \eta_0e^{-2\sqrt{\bar{N}}\gamma_{0}t_d},
\end{equation}
where $\gamma_{0}$ is the dephasing rate at single-photon transduction. The experimental results we present later align well with this formula and reveal two counterintuitive outcomes: the storage efficiency at the single-photon level is significantly higher than that in the multi-photon case, which stands in sharp contrast to the difficulty of achieving high efficiency with single photons in storage experiments~\cite{Wang2019}. On the other hand, this formula holds not only when $\bar{N} \ge 1$ but also when $\bar{N}>0.1$, which may stem from the fact that in the range of $0.1 < N < 1$, there is still a certain probability of multi-photon events occurring.

Fig.~\ref{fig:4}c shows the dependence of measured $\eta_\mathrm{I}$ on the input MW photon numbers. The measurements align well with the fitting results to Eq.~(\ref{Efficiency2}) with an intrinsic efficiency $\eta_0 = 88\%$ and single-photon dephasing rate $\gamma_{0}/2\pi$ = 12.8 kHz, which is close to the parameter-free theoretical prediction $\gamma_{0}/2\pi=10.8$ kHz by using the mean-field model(see Supplementary Note~5). This result illustrates that the model is reasonable and our OMQT maintains a high efficiency around 90$\%$ in the regime of single-photon-level transduction. Finally, we vary the probe beam power in order to regulate the OD of MW transition. In the absence of Rydberg decoherence ($\gamma_{51}$ = 0), it is shown that the ideal $\eta_\mathrm{I}$ can be close to unity with the scaling law of optimal storage~\cite{Gorshkov2007a}: $\eta_\mathrm{I} \approx 1 - \beta/d_{M}$. The measured efficiency as a function of $d_{M}$ is shown in Fig.~\ref{fig:4}d. The $\eta_\mathrm{I}$ initially grows with $d_{M}$ but becomes saturated around $d_{M}$ = 10$^{6}$. The data generally agree with the scaling law of optimal storage, implying that the OMQT benefits from the ultra-high OD and backward retrieval, and thus approaches the optimal storage of MW photons. In the region of $d_{M} \geq$ 10$^{6}$, the deviation between the measured values and the optimal storage may arise from the increased Rydberg dephasing owing to a higher atom density populated in the Rydberg state $|3\rangle$. However, the increased van der Waals interaction between levels $|3\rangle$ and $|5\rangle$ enhances spin-wave decoherence, resulting in efficiency degradation at higher ODs(see Supplementary Note~5).

\section{Discussion }

The direct MO transduction method~\cite{Han2018, Tu2022, Kumar2023, Borowka2024} relies on a steady-state process that requires impedance matching conditions to achieve complete conversion. In contrast, our OMQT employs a distinct mechanism: through adiabatic storage operations with time-dependent modulation and non-steady-state dynamics, we suppress the influence of dark states~\cite{Han2018, Tu2022}, thereby eliminating their detrimental effects on transduction efficiency. Moreover, the OMQT incorporates quantum memory-optimized control strategies~\cite{Gorshkov2007a, Wang2019} and exhibits a parameter-dependent efficiency profile fundamentally different from direct MO transduction. The exceptionally high OD for MW photons enables near-unitary efficiency, which raises the theoretical upper limit of conversion efficiency significantly beyond that achievable with DC approaches. In addition, based on a different mechanism, the phase-mismatching mainly degrades the coherence time of spin waves while having little effect on the transduction efficiency.

Beyond these, our OMQT offers additional advantages for practical application. First, it integrates quantum storage and MO conversion functions, creating a unique device for synchronization and storage essential for generating entanglement between distant nodes in a quantum network. Second, the OMQT requires only a simple optical filter to achieve MO conversion for the incident single-photon energy, whereas conventional ambient MO transduction methods require an average photon number of $\bar{N}>50$ to observe detectable conversion~\cite{Tu2022} due to their limited signal-to-noise ratio. Finally, the OMQT exhibits a broader bandwidth, as the system does not necessitate reaching a steady state.

In summary, we have theoretically proposed and experimentally demonstrated a Rydberg atom-based MO transducer enhanced by an absorptive quantum memory. Our methods can be extended to various types of MO transducer systems~\cite{LambertReview2019, LaukReview2020, XHan2021}. To extend the storage time, an additional transition can be employed immediately after the write process to transfer the collective excitations from the Rydberg state $|5\rangle$ to a hyperfine clock state. This approach enables a storage time on the order of seconds by eliminating the residual spin-wave momentum~\cite{Li2016}. A more challenging extension involves realizing our OMQT using Rydberg atoms trapped in superconducting circuits~\cite{KiffnerNJP2016, Axline2018, FortaghPRR2022, Kumar2023}, especially inside the millikelvin cryogenic environment that is suitable for superconducting qubits integration~\cite{LandraPRA2019}. In a cryogenic environment, the MOT-cooled atoms should be transferred into a magnetic~\cite{FortaghNC2017} or optical~\cite{FortaghPRapp2025} trap near the waveguide to mitigate atomic-motion-induced decoherence. This quantum-memory-enhanced transducer can also provide an additional degree of freedom for detection applications, such as radio astronomy~\cite{Komatsu2022} and next-generation MW sensors~\cite{Schlossberger2024} in the single-photon domain.

\section{Methods}

\subsection{Entanglement generation between remote solid-state qubits}

We assume that the solid-state qubit connected to the waveguide can generate a MW time-bin qubit at a rate~\cite{FortaghNC2017, KurpiersPRApp2019} of $r_{\text{ss-wg}}$ = 1 MHz. After the photonic conversion within a given noise photon $n_{\text{p}}$, the success probability is $P = 1 / (1 + n_{\text{p}})$. For the DC case, the total time per attempt is expressed as $\tau_{\text{total}}^{\text{DC}} = \tau_{\text{emi}} + \tau_{\text{con}}^{\text{DC}} + \tau_{\text{BSM}}$, where $\tau_{\text{emi}}$ represents the period of MW emission, $\tau_{\text{con}}^{\text{DC}} = 1 \mu s$ is the direct conversion time, and $\tau_{\text{BSM}}$ denotes the time for Bell-state measurement. Thus, we calculate the entanglement generation rate between remote solid-state qubits as
\begin{equation}
\label{EntanglementRate}
R_{\text{DC}} = \frac{1}{2} r \eta_{\text{DC}}^2 \eta_{\text{fib}}^2 \eta_{\text{det}}^2 P^2 ,
\end{equation}
where $r = \min\left( 1 / \tau_{\text{total}}^{\text{DC}},\ r_{\text{ss-wg}} \right)$ is the entanglement attempt rate in DC scheme, $\eta_{\text{fib}}$ and $\eta_{\text{det}}$ are the fiber transmission and detector efficiencies, respectively, and the factor $1/2$ accounts for the success probability of the Bell state measurement.

In the OMQT scheme, the role of multiplexed quantum memory is to store the solid-state qubits such that the operations can be timed appropriately, thereby enhancing the quantum communication between remote solid-state systems. The success possibility per optical transmission is $p = P \cdot \eta_{\text{QM}} \cdot \eta_{\text{fib}} \cdot \eta_{\text{det}}$. The microwave attempt rate is $r_{\text{s-e}} = 1 / (\tau_{\text{emi}} + \tau_{\text{abs}})$, where $\tau_{\text{emi}}$ and $\tau_{\text{abs}}$ denote the emission and absorption times of the microwave photon in memory, respectively. The average time to establish a microwave-optical entanglement link, $\tau_{\text{s-e}}$, accounting for the probabilistic success, can be expressed as:
\[
\tau_{\text{s-e}} = \frac{1}{r_{\text{s-e}}} \left[ p^2 + \sum_{i=0}^{\infty} (i+2)p^2 q^{i+1} \left( q^{i+1} + 2 \sum_{j=0}^{i} q^j \right) \right],
\]
where $q = 1-p$, $p^2$ denotes the success probability of first trial, and the following term of $p^2 q^{n-1} \left( q^{n-1} + 2 \sum_{j=0}^{n-2} q^j \right)$ represents that of the $n$th trial~\cite{Kurokawa2022}. The total operation time is $\tau_{\text{total}}^{\text{QM}} = \tau_{\text{s-e}} + \tau_{\text{readout}} + \tau_{\text{BSM}}$, which sums the times for microwave emission, microwave-optical entanglement establishment, optical readout after storage, and Bell state measurement. The storage time $\tau_{\text{coh}}$ restricts the maximum number of attempts to be $N_{\max} = \tau_{\text{coh}} / (\tau_{\text{emi}} + \tau_{\text{abs}})$, and thus, in the case of establishment time $\tau_{\text{s-e}} > \tau_{\text{coh}}$, a modified success probability $p^{\prime} = 1 - q^{N_{\max}}$ is used to determine $\tau_{\text{s-e}}$. Finally, the entanglement generation rate is given by $R_{\text{QM}} = \min\left( 1 / \tau_{\text{total}}^{\text{QM}},\ r_{\text{ss-wg}} \right)$. We plot the curves of Figure~\ref{fig:1}c assuming $\eta_{\text{det}}$ = $\eta_{\text{fib}}$ = 0.9, $\tau_{\text{emi}}$ = $\tau_{\text{abs}}$ = 1 $\mu$s, $\tau_{\text{readout}}$ = 0.1 $\mu$s, and $\tau_{\text{BSM}} = 10\,\text{ns}$.

\subsection{Spin-wave decoherence induced by atomic random motion}

After MW photons are stored as a spin wave, the atomic random motion perturbs the spin wave phase, reducing the probability of successful optical conversion over time. The decoherence timescale is determined by the average time for the atoms to traverse $1/2\pi$ of the spin-wave wavelength $\lambda_{\mathrm{sw}}$,
\begin{equation}
\tau_{\mathrm{sw}} = \frac{\lambda_{\mathrm{sw}} } {2 \pi u},
\end{equation}
where $u=\sqrt{k_{\mathrm{B}} T / m}$ is the one-dimensional mean thermal speed ($k_{\mathrm{B}}$: Boltzmann constant; $m$: atomic mass). For our collinear optical layout, the wavelength $\lambda_{\mathrm{sw}}=\left|\mathbf{k}_{\mathrm{P}}-\mathbf{k}_{\mathrm{A}}\right|\simeq 1.3 \mu m$, yielding $\tau_{\mathrm{sw}} = 1.7 \mu s$ at $T$ = $150 \mu K$. In Fig.~\ref{fig:3}c, since $\sqrt{\bar{N}}\gamma_{0} \ll 1/\tau_{sw}$ for $\bar{N}$ = 0.1, the dispersion of Rydberg level shifts dominates the efficiency but negligibly affects storage time. The fitted storage time ($0.9 \mu s$) is the same order of timescale of the Rydberg spin-wave decoherence.

\section{Data Availability}

The data that support the findings of this study are available within the article and its Supplementary Information. Source data for the figures are provided with the paper. Any additional data can be obtained from the corresponding author upon request.


\section{Acknowledgements}

We thank Y.F. Mei and C.L. Zou for useful discussions.

\section{Funding Statement}

The work was supported by the Quantum Science and Technology-National Science and Technology Major Project (Grant No. 2021ZD0301705), the National Natural Science Foundation of China (Grants Nos. 12474488, 12225405, and 12534011), the National Key Research and Development Program of China (Grants Nos. 2021YFA1402004 and 2022YFA1405300), Guangdong Provincial Quantum Science Strategic Initiative (Grants Nos. GDZX2405004 and GDZX2304002), and the Guangdong Basic and Applied Basic Research Foundation (Grants Nos. 2025B1515020075 and 2023A1515030003).

\section{Author Contributions}

K.Y.L., H.T.T., H.Y., and S.L.Z. designed the experiment. H.T.T., S.Y.Q., X.H.L., Y.Q.G., and Z.Q.D carried out the experiments. H.T.T., K.Y.L., S.Y.Q., X.D.Z. and Y.X. conducted raw data analysis. H.T.T., K.Y.L., H.Y. and S.L.Z. wrote the paper, and all authors discussed the paper contents. K.Y.L., H.Y. and S.L.Z. supervised the project.

\section{Competing Interests}

The authors declare no competing interests.

\clearpage
\begin{table}[htbp]
\centering
\caption{Relevant frequencies, polarizations, waists, and electric dipole moments $|\boldsymbol{\mu}_{l m}|$ for the six fields.}
\label{tab:fields}
\begin{tabular}{|l|c|c|c|c|}
\hline
Field & $f$ or $\lambda$ & Polarization & Waist ($\mu$m) & $|\boldsymbol{\mu}_{l m}|$ ($ea_{0}$) \\
\hline
$\Omega_{\mathrm{P}}$ & $\sim$384.2 THz & $\sigma^{+}$ & 66$\pm$3 & 2.99  \\
$\Omega_{\mathrm{L}}$ & 780.2 nm & $\sigma^{-}$ & NA & 1.22  \\
\hline
$\Omega_{\mathrm{A}}$ & $\sim$624.1 THz & $\sigma^{-}$ & 128$\pm$3 & 0.0089 \\
$\Omega_{\mathrm{R}}$ & 479.7 nm & $\sigma^{+}$ & 128$\pm$3 & 0.0098  \\
\hline
$\Omega_{\mathrm{M}}$ & 37.5 GHz & $\sigma^{+}$ & NA & 1271  \\
\hline
$\Omega_{\mathrm{W}}$ & 22.1 GHz & $\sigma^{-}$ & NA & 541 \\
\hline
\end{tabular}
\end{table}

\clearpage

\clearpage
\begin{figure}[p]
\begin{center}
\includegraphics[width=13.0cm]{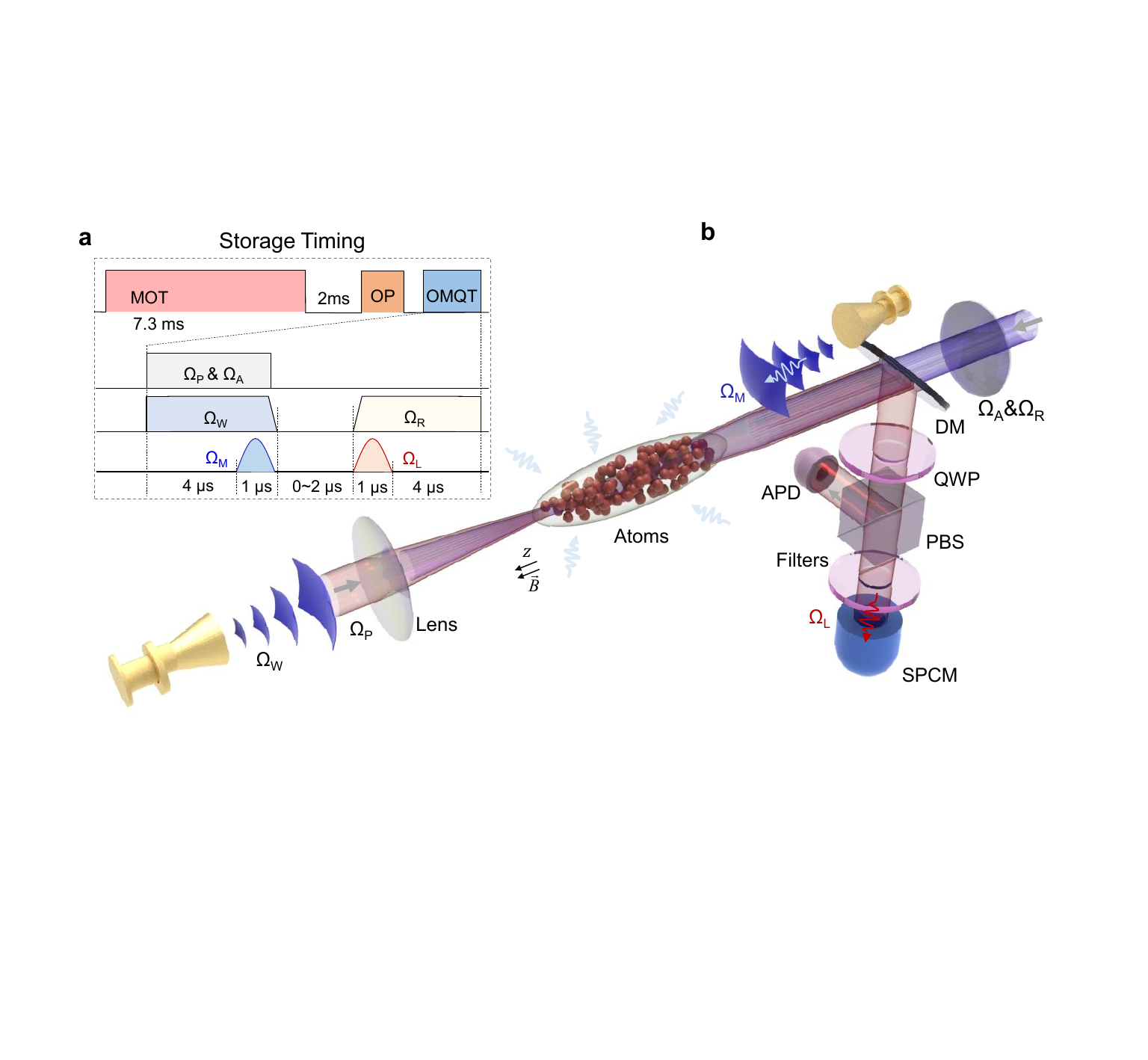}
\caption{\label{fig:2} \textbf{Proof-of-concept OMQT in cold atoms.} \textbf{a}, Schematic of the time sequence containing MOT loading, optical pumping (OP), and OMQT, which comprises waveforms for the input and recalled pulses ($\Omega_{\rm{M}}$ and $\Omega_{\rm{L}}$), the write and read pulses ($\Omega_{\rm{W}}$ and $\Omega_{\rm{R}}$), and two auxiliary fields ($\Omega_{\rm{P}}$ and $\Omega_{\rm{A}}$). The energy levels and the related six fields are illustrated in the inset of Fig.~1a. \textbf{b}, Experimental setup, including a cigar-shaped atomic cloud, antennas, lenses, a dichroic mirror (DM), a quarter-wave plate (QWP), a polarization beam splitter (PBS), filters, an avalanche photodiode (APD), and a single-photon counting module (SPCM).}
\end{center}
\end{figure}

\clearpage
\begin{figure}[p]
\begin{center}
\includegraphics[width=15.0cm]{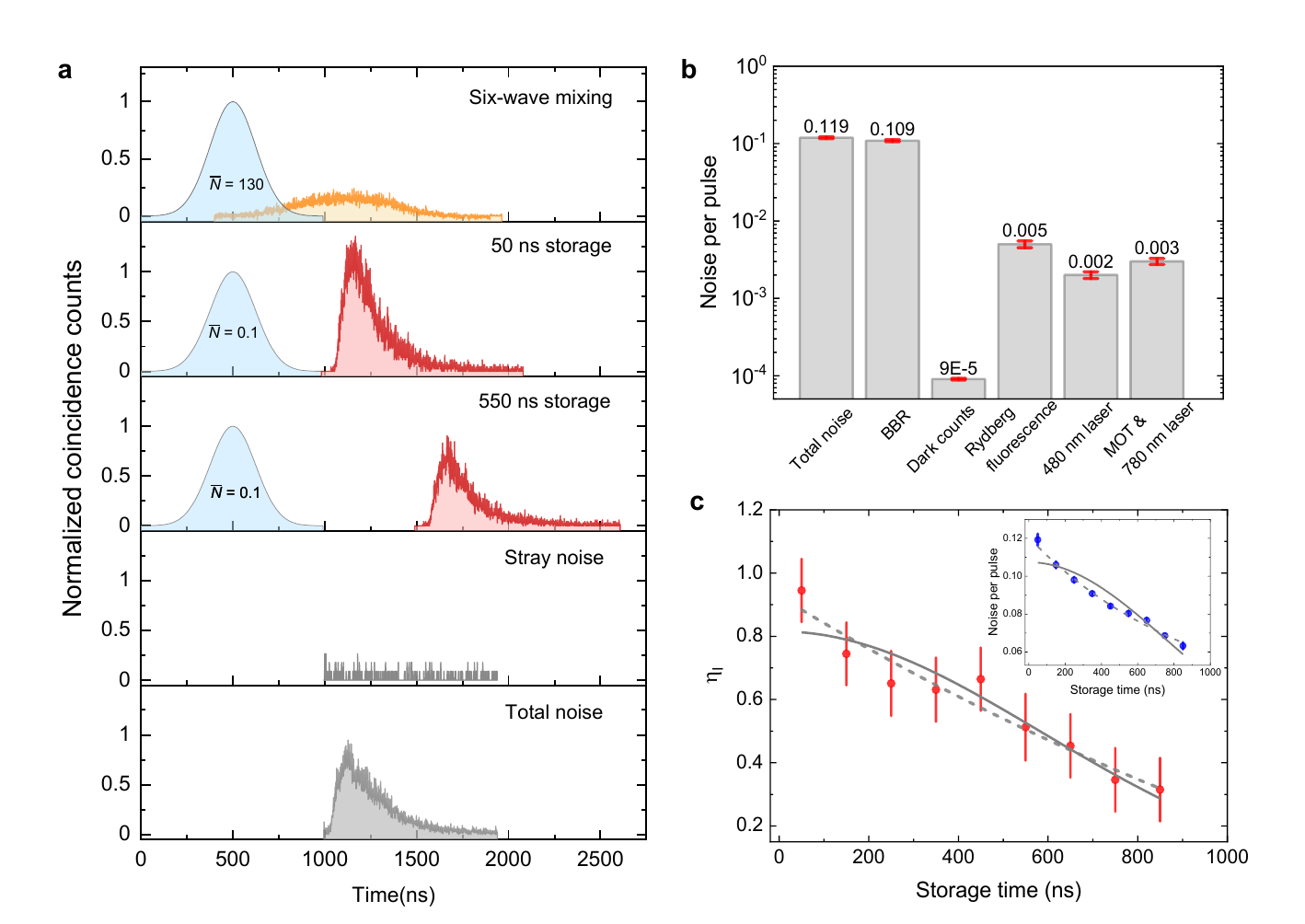}
\caption{\label{fig:3} \textbf{On-demand MO transduction and its noise characteristics.} \textbf{a}, Temporal waveforms of input MWs, slow-light pulses, retrieved optical pulses, and recalled noises throughout the conversion window. \textbf{b}, Registered noise components for each pulse are outlined after a 50 ns storage time. \textbf{c}, ASE $\eta_\mathrm{I}$ as a function of storage time for $\bar N=0.1$. The inset in \textbf{c} illustrates the counts of recalled noise photons as a function of storage time. The count data were collected over $2\times 10^4$ cycles. The solid (dashed) line represents fitting results derived from the Gaussian (exponential) decay function. The error bars represent the standard deviation from three measurements. The experimental parameters $\{\Omega_\mathrm{P}, \Omega_\mathrm{A}, \Omega_\mathrm{W}, \Omega_\mathrm{R}, \Gamma_{2}, \gamma_{3}, \Gamma_{4}, \Gamma_{6}\}$ = $2\pi \{2.1, 7.6, 1.8, 9.0, 6.0, 0.5, 0.001, 1\}$ MHz.}
\end{center}
\end{figure}

\clearpage
\begin{figure}[p]
\begin{center}
\includegraphics[width=15cm]{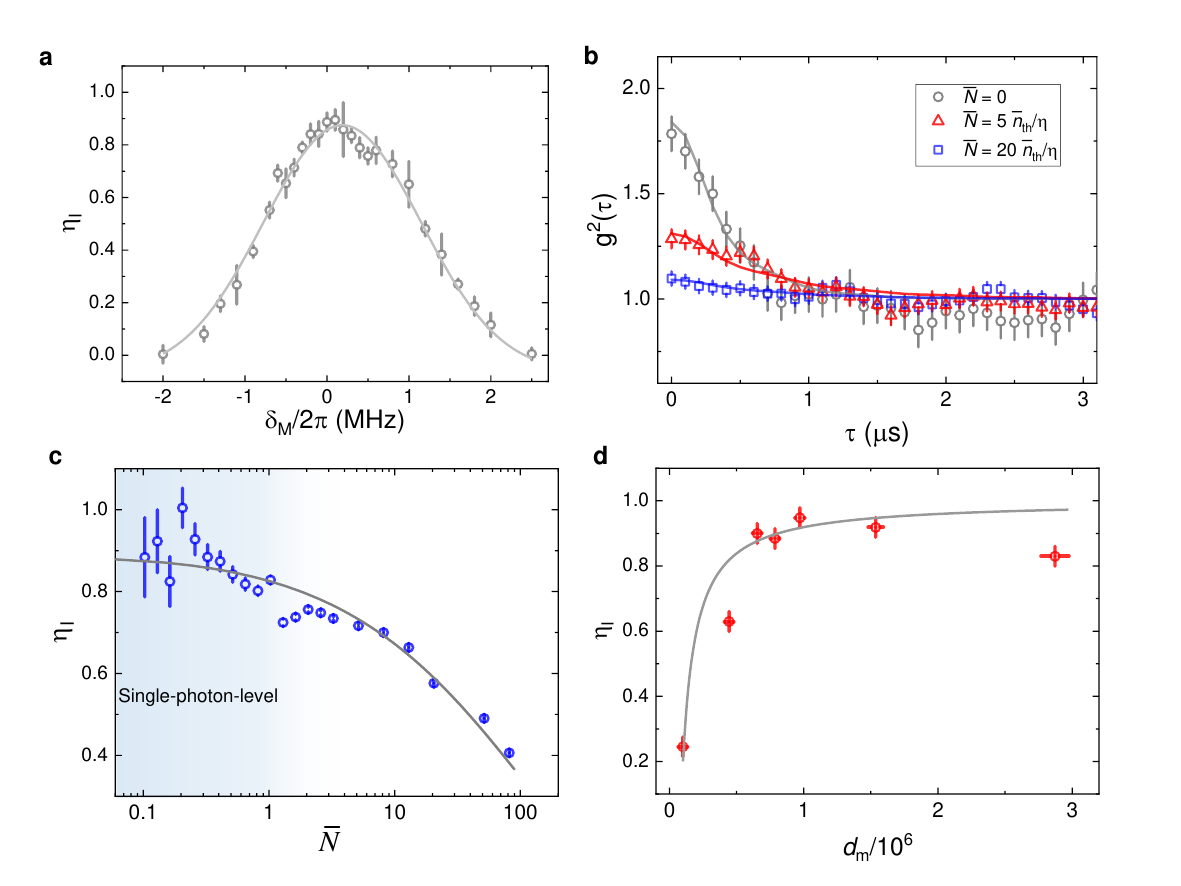}
\caption{\label{fig:4} \textbf{Storage efficiency and intensity autocorrelations.} \textbf{a}, ASE $\eta_\mathrm{I}$ versus $\delta_M$. The data were collected at $\bar{N} = 0.3$, and the data points are fitted using a Gaussian function with an offset. \textbf{b}, Second-order autocorrelation functions of retrieved optical photons at $\bar N = 0, 5 \bar n_{th}/\eta_\mathrm{I}, 20 \bar n_{th}/\eta_\mathrm{I}$. The solid curves represent theoretical predictions without the free parameters, while the error bars represent the standard deviation from three measurements. \textbf{c}, $\eta_\mathrm{I}$ as a function of $\bar{N}$ at a 50 ns storage time. The shaded area indicates the single-photon-level transduction. The solid curve is obtained by fitting the data ($\bar{N}>0.1$) to Eq.(4), yielding the fitting results of $\eta_0 = 88\%$ and $\gamma_{0}/2\pi = 12.8$ kHz. \textbf{d}, $\eta_\mathrm{I}$ versus the OD $d_{\text{M}}$ with $\bar{N} = 1$. The solid line illustrates the fitted scaling function for optimal storage, $\eta_\mathrm{I} = 1 - 80796/d_{\text{M}}$. The $d_{\text{M}}$ error bars indicate the standard deviation from sixty measurements, and the ASE uncertainties in \textbf{a}, \textbf{c}, and \textbf{d} are computed similarly to those in Fig.~3c.}
\end{center}
\end{figure}

\let\addcontentsline\oldaddcontentsline

\clearpage

\begingroup   

\setcounter{equation}{0}
\setcounter{figure}{0}
\setcounter{table}{0}
\renewcommand{\theequation}{\arabic{equation}}
\renewcommand{\thefigure}{\arabic{figure}}
\renewcommand{\thetable}{\arabic{table}}

\renewcommand{\figurename}{Supplementary Figure}
\renewcommand{\tablename}{Supplementary Table}


\begin{center}
  {\fontsize{16pt}{20pt}\selectfont\bfseries  Supplementary Materials for\\ Quantum-memory-assisted on-demand microwave-optical transduction}
\end{center}

\renewcommand{\theequation}{\arabic{equation}}
\setcounter{equation}{0}
\renewcommand{\thefigure}{\arabic{figure}}
\setcounter{figure}{0}
\maketitle

\section* {Supplementary Note 1 -- Transduction mechanism}

  As shown in Figure 1a of main text, we utilize a cascade-type five-level system to store the MW photons in waveguide mode, and then on-demand retrieve the optical photons in another cascade-type EIT system. In the writing scheme, the atomic-level system involves three Rydberg states, and in general sustains the dipole-mediated interactions between atoms excited into Rydberg manifolds. Because the atomic density is not high (i.e., less than one atom within a blockade sphere) in a medium driven by the auxiliary fields $|\Omega _P| \ll |\Omega _A|$, we can ignore the photon blockade and utilize the effective dephasing terms to phenomenologically quantify the transmission property in the storage model. Assuming that the input MW is sufficiently weak, the writing process is regarded as a perturbation of the three-level atomic system $|1\rangle$, $|2\rangle$, and $|3\rangle$. Most atoms are coherently trapped in the dark state of the three-level system, namely the zero-order solution for atomic coherences\cite{Hsiao2018}. Since the dephasing rate of state $|3\rangle$ is small compared with the auxiliary fields $\Omega_P$ and $\Omega_A$, we find the zero-order atomic density operators as follows:
\begin{equation}
\rho_{11}=\frac{\left|\Omega _A\right|{}^2}{\left|\Omega _P\right|{}^2+|\Omega _A|^2},
\end{equation}
\begin{equation}
\rho_{33}=\frac{\left|\Omega _P\right|{}^2}{\left|\Omega _P\right|{}^2+|\Omega _A|^2},
\end{equation}
\begin{equation}
\rho_{13}=-\frac{\Omega _P \Omega _A}{\left|\Omega _P\right|{}^2+|\Omega _A|^2}.
\end{equation}
Under the slowly-varying envelope approximation, we analyze the storage and retrieval dynamics of an uniform ensemble along the coplanar waveguide (z direction) by solving the first-order Maxwell-Bloch equations:
\begin{equation}
\left(  \partial _ { z } + \frac{1}{c} \partial _ { t } \right) \Omega_M^{+}  = \frac{-i \rho_{13} } {2\rho_{11}} F  n_3 \sigma_{\rm{M}} \Gamma_{4} P_{41}^{+},
\end{equation}
\begin{equation}
\left(  \partial _ { z } - \frac{1}{c} \partial _ { t } \right) \Omega_L^{-}  = \frac{-i } {2}  n_1 \sigma_{\rm{L}} \Gamma_{6} P_{61}^{+},
\end{equation}
\begin{equation}
\partial_{ t } P_{41}^{+}=-\gamma_{41} P_{41}^{+}+  \frac{i}{2} \rho_{13}\Omega_M^{+} + \frac{i}{2} \Omega_W^{-} P_{51},
\end{equation}
\begin{equation}
\partial_{ t } P_{61}^{-} =-\gamma_{61} P_{61}^{-}+  \frac{i}{2} \rho_{11}\Omega_L^{-} + \frac{i}{2} \Omega_R^{+} P_{51},
\end{equation}
\begin{equation}
\partial_{ t }  P_{51} =-\gamma_{51}  P_{51}+  \frac{i}{2} (\Omega_W^{-})^{*} P_{41}^{+} + \frac{i}{2} (\Omega_R^{+})^{*} P_{61}^{-}.
\end{equation}
Here the superscript $+$ ($-$) indicates the fields or atomic coherence in the forward propagation (backward retrieval) direction, and the collective excitation $P_{51}$ represents the spin wave component of Rydberg polaritons when MW photons are mapped inside the ensemble.

The input MW pulse has a Gaussian wavepacket with a FWHM duration of $T_{p}$, giving
\begin{equation}
\Omega_M(t, z=0)=\Omega_{M 0} \exp \left(-2 \ln 2 \frac{t^2}{T_{p}^2}\right).
\end{equation}
In the frequency domain, the input pulse takes the form as
\begin{equation}
W_M(\omega, z=0)=\frac{\Omega_{M 0} T_{p}}{\sqrt{4 \ln 2}} \exp \left(-\frac{\omega^2 T_{p}^2}{8 \ln 2}\right).
\end{equation}
Due to the short storage time and high OD, we only  need to consider the EIT slow-pulse transmission in the overall storage efficiency analysis. Since both auxiliary fields, $\Omega_W$ and $\Omega_R$, are significantly stronger than $\Omega_M$ in the respective cascade-EIT schemes, their variations are negligible  and can be treated as the constants. Then we take the Fourier transform on the atomic coherence $P_{j1}$ and two field envelopes ($\Omega_M$ and $\Omega_L$) to the frequency domain, for instance $R_{j1}=1 / \sqrt{2 \pi} \int_{-\infty}^{\infty} P_{j1} e^{i \omega t} d t$. Supplementary Equations (4)--(8) are written as:
\begin{equation}
\frac{\partial W_M}{\partial z}-\frac{i\omega}{c}  W_M=\frac{-i \rho_{13} } {2\rho_{11}} F n_3 \sigma_{\rm{M}} \Gamma_{4} R_{41},
\end{equation}
\begin{equation}
\frac{\partial W_L}{\partial z}+\frac{i\omega}{c}  W_L= \frac{-i } {2}  n_1 \sigma_{\rm{L}} \Gamma_{6} R_{61},
\end{equation}
\begin{equation}
-i \omega R_{41}= =-\gamma_{41} R_{41}+  \frac{i}{2} \rho_{13}W_M + \frac{i}{2} \Omega_W R_{51},
\end{equation}
\begin{equation}
-i \omega  R_{61} =-\gamma_{61} R_{61}+  \frac{i}{2} \rho_{11}W_L + \frac{i}{2} \Omega_R R_{51},
\end{equation}
\begin{equation}
-i \omega  R_{51} =-\gamma_{51}  R_{51}+  \frac{i}{2} (\Omega_W)^{*} R_{41} + \frac{i}{2} (\Omega_R)^{*} R_{61}.
\end{equation}
By solving Supplementary Equations (11), (13), and (15), we find

\begin{equation}
\begin{split}
W_M(\omega, z)=W_M(\omega, 0) \exp [\frac{i \omega z}{c} +\frac{-i |\rho_{13}|^2 F n_3 \sigma_{\rm{M}} z \Gamma_{4} } {4\rho_{11}} \frac{i\omega-\gamma_{51}}{\left(i\omega-\gamma_{41}\right)\left(i\omega-\gamma_{51}\right)+\Omega_W^2 / 4}].
\end{split}
\end{equation}

Substituting the expression of $W_M(\omega, 0)$ into Supplementary Equation (16) and making an inverse Fourier transform, we get the solution of the MW pulse propagating through atomic ensemble as

\begin{equation}
\begin{split}
\Omega _M(t,z=L)=\frac{\Omega _{M0} T_{p}}{ \sqrt{8 \pi\ln2 }}\int _{-\infty }^{\infty }\text{d$\omega $}
\exp \{(-i t +\frac{i L }{c})\omega -\frac{T_{p}^2 \omega ^2}{8\ln2 } +\frac{\rho_{33} F d_{M} \Gamma _4  \left(i \omega -\gamma _{51}\right)}{4 \left(i \omega -\gamma _{41}\right) \left(i \omega -\gamma _{51}\right)+\Omega _W^2}\},
\end{split}
\end{equation}

where $d_{M} = n_3 \sigma_M L$  is the optical depth of MW transition $|3\rangle\leftrightarrow|4\rangle$. Assuming that $\Omega _W^2>>\gamma _{41}\gamma _{51}$ and truncating the susceptibility function to the $\omega^2$ term allow us to reduce the above integral result as:
\begin{small}
\begin{equation}
\Omega _M(t,z=L)=\frac{\Omega _{\text{M0}}}{\zeta_M } \exp \left[-\frac{ \gamma _{51} \Gamma _4\rho_{33} F d_{M}}{\Omega _W^2}-2 \ln 2  \left(\frac{t-t_{dM}}{\zeta_M  T_{p}}\right)^2\right ].
\end{equation}
\end{small}
In the above equation, we introduce  a factor,
\begin{equation}
\zeta_M  = \sqrt{\text{1+} \frac{32 \ln2\gamma _{41} \Gamma _4\rho_{33} F d_{M}}{T_{p}^2 \Omega _W^4}},
\end{equation}
to characterize the amplitude attenuation and temporal broadening of the MW pulse after propagation through the EIT medium. The group delay time is given by
\begin{equation}
t_{dM}=\frac{L}{c}+\frac{\Gamma _4 \rho_{33} F d_{M}}{\Omega _W^2} \simeq \frac{\Gamma _4 \rho_{33} F d_{M}}{\Omega _W^2},
\end{equation}
since the group velocity is far less than the light speed. Integrating Supplementary Equation (18) over the time, this leads to the MW transmission efficiency,
\begin{equation}
\eta_M=\frac{e^{-2 \gamma _{51} t_{dM}}}{\sqrt{1+\alpha_{M}/{F d_{M}}}},
\end{equation}
where we have defined
\begin{equation}
\alpha_{M}  = 32\ln2 \frac{ \gamma _{41} t_{dM}^2}{ \Gamma _4 T_{p}^2 },
\end{equation}
and $\gamma _{41}= \Gamma _4$, $T_P=500 ns$.

The same treatment can be applied for the stage of the optical photon retrieval. The ground state $|1\rangle$ and Rydberg state $|5\rangle$ are the common energy levels of writing and reading systems, and hence, in this case, the Rydberg dephasing $\gamma _{51}$ dominates the optical transmission efficiency as indicated by Supplementary Equation (8). By solving Supplementary Equations (12), (14), and (15), one obtain the optical transmission efficiency with a optical delay time $t_{dL}$, giving
\begin{equation}
\eta_L=\frac{e^{-2 \gamma _{51} t_{dL}}}{\sqrt{1+\alpha_{L}/{d_{L}}}},
\end{equation}
where
\begin{equation}
\alpha_{L}  = 32\ln2 \frac{ \gamma _{61} t_{dL}^2}{ \Gamma _6 T_{pL}^2 }.
\end{equation}
where $T_{pL}$ = 620 ns is the FWHM of slow light delay pulse in Fig. 3a of main text, and $\gamma_{61}=\Gamma_6/2$. Together, the transmission efficiency is the product of these two,
\begin{equation}
\eta_t =\eta_M \eta_L =\frac{e^{-2 \gamma_{51} t_d}}{\sqrt{\left(1+\alpha_{M}/{F d_{M}}\right)\left(1+\alpha_{L}/{d_{L}}\right)}}.
\end{equation}
As shown in the middle panel of Fig. 3a in main text, $t_d \simeq$  623 ns is the measured total delay time with $t_{dM}\approx$ 500 ns and $t_{dL}\approx$ 123 ns. By setting a unity filling factor ($F$ = 1), we obtain an area-normalized efficiency $\eta_\mathrm{I} \approx \eta_t \simeq 0.9 $ in the free-space case, considering the relevant parameters: $\gamma_{51}/2\pi$ = 12.8 kHz, $d_L$ = 141$\rho_{11}  \approx$ 122, $d_M= 7.5\times 10^5$, \(\alpha_M \simeq 61.61\), and \(\alpha_L \simeq 0.44\).

In our experiment, the area-normalized efficiency $\eta_\mathrm{I}$ is determined as the the ratio of retrieved optical photons $N_\mathrm{L}$ to MW photons $\bar N$ incident on the medium~\cite{Tu2022,JHan2018,Vogt2019,Borowka2024},
\begin{equation}
\eta_\mathrm{I}=\frac{N_\mathrm{L}}{\bar N}=\frac{ \hbar \omega_\mathrm{M} (C_\mathrm{L}-C_\mathrm{N}) }{\kappa A_\mathrm{L} \int I_\mathrm{M}(t) \mathrm{d} t  },
\end{equation}
where $\hbar$ is the reduced Planck constant, $\omega_\mathrm{M}$ is the angular frequency of MW photon, $C_\mathrm{L}$ ($C_\mathrm{N}$) is the coincidence count measured in the presence (absence) of input MW, $A_\mathrm{L}$ is the receiving size of atomic medium (see Supplementary Note 5), $\kappa$ is the detection efficiency (39\%) along the optical path, and $I_\mathrm{M}$ is the MW intensity. 

The $\eta_\mathrm{I}$ is an ideal quantity when the interaction region is much larger than the wavelength of the MW, but caution is needed if the interaction region is substantially subwavelength for the MW fields\cite{Borowka2024}. However, this quantity remains a practical physical metric for characterizing the conversion efficiency in free-space configurations, even under the latter condition. As experimentally demonstrated in Ref.\cite{Borowka2024}, the measured area-normalized efficiency in their system agrees with theoretical expectations. Similarly, in our experimental setup, the measured area-normalized efficiency aligns closely with the theoretical derivation detailed in Supplementary Note 1. On the other hand, we theoretically demonstrate that the number of optical photons converted by microwave background radiation per pulse is approximately 0.08 (0.109) when the $\eta_\mathrm{I}$ is 91\% (93\%), respectively (see Supplementary Note 6). This value aligns with the observed thermal noise count of 0.109 per pulse, providing strong independent evidence that the $\eta_\mathrm{I}$ can serve as a reliable metric for measuring conversion efficiency.

\section*{Supplementary Note 2 -- Experimental setup}
A cigar-shaped $^{87}$Rb ensemble with a volume of 4 $\times$ 4 $\times$ 20 mm$^{3}$ serves as the MO transducer, and the atomic number density $ n_{a t} \approx 2.4 \times$ 10$^{\rm{10}}$ cm$^{-3}$. { For the magneto-optical trap, each cooling laser beam has a power of 25 mW and a radius of 1.6 cm. The total power of two repumping laser beams is around 35 mW with the same radius as the cooling beam.} The 780 nm laser (auxiliary field $\Omega_P$) and the seeds of two 480 nm lasers (fields $\Omega_A$ and $\Omega_R$) are frequency-locked to a high-finesse Fabry-Perot cavity by the Pound-Drever-Hall technique\cite{Tu2022, Liao2020, JHan2018, Vogt2019, XHLiu2022}. The MW fields $\Omega_M$ and $\Omega_W$, derived from two commercial RF generators, are locked to a rubidium clock, and the input near-Gaussian MW pulse is produced by using an arbitrary waveform generator and a double balanced mixer, as described in Supplementary Note 3. The MOT is positioned in the far-field region of a helical antenna that has a gain of 20 dB, a horn size of 34 mm, and a 30 cm spacing between the antenna and atomic ensemble. Consequently, the MW can be approximated as a plane wave exhibiting a highly uniform and stable electric field.

In order to isolate the stray photons from single-photon detection, we use a narrow-band-pass filter at 780 nm to filter out the 480 nm light, and use a polymer film linear polarizer to isolate the 780 nm copropagating laser beam. Despite a narrow frequency difference of 0.27 GHz between the recalled photons and the 780 nm auxiliary beam, our transducer benefits from its inherent storage capability, demonstrating significant temporal isolation against stray photons. The optical setup possesses the spatial isolation due to single-mode optical fiber, and consequently reaches an overall isolation of 50 dB. Taking into account the optical component transmission (91\%), filter transmission (80\%), fiber coupling efficiency (83\%), and SPCM detection efficiency (65\%), we estimate the photon number retrieved from the atomic ensemble. { We use a laser beam to measure the collection probability of photons from the atomic ensemble to the fiber output. Taking into account the SPCM quantum efficiency, we obtain the optical detection efficiency (39\%).  } The overall detection efficiency along the optical path can be improved by replacing current components with higher-quality optical devices and superconducting nanowire single-photon detectors, which typically have a quantum efficiency exceeding 90\% in the near-infrared band. 

\section*{Supplementary Note 3 -- Generation of a Gaussian microwave pulse and the photon number calibration}

\begin{figure*}
\begin{center}
\includegraphics[width=14cm]{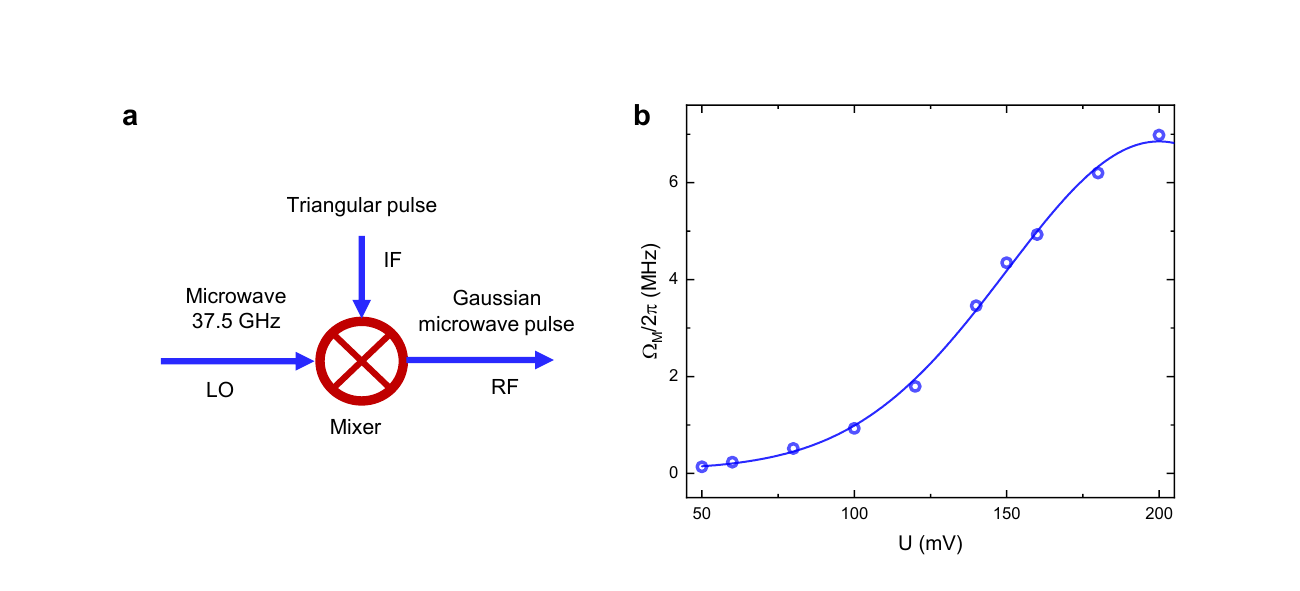}
\caption{\label{supfig:1}{\textbf{Generation of a Gaussian MW pulse.} \textbf{a}, Schematic diagram of the Gaussian MW pulse generation.  \textbf{b}, RF output amplitude as a function of the input voltage of IF port.  The RF amplitude is proportional to the microwave field Rabi frequency $\Omega_M$, and calibrated by the cold Rydberg atomic Microwave electrometry. The circle points denote the result of experimental measurements, and the solid line is a Gaussian fit.  }
}
\end{center}
\end{figure*}

For the EIT-based quantum memory, the signal photons can be efficiently stored in an atomic ensemble when the input pulse has a Gaussian temporal waveform\cite{Wang2019}. Here we generate the near-Gaussian MW pulses by using a RF mixer (Marki MM11044L), as shown in Supplementary Figure \ref{supfig:1}a. A continuous microwave of 37.5 GHz serves as the local oscillator (LO) for the mixer, while a triangular pulse generated by an arbitrary wave generator (Agilent 81150A) is applied to the IF port to control the pulse envelope. To determine the dependence of MW Rabi frequency \( \Omega_M \) on the input voltage \( U \) at the IF port, we vary the IF voltage point by point and measure the MW amplitude $|E|=\hbar \Omega_{\mathrm{M}} /|\boldsymbol{d}_{34}|$ using the EIT-ATS method\cite{Liao2020}. The measurement results are shown in Supplementary Figure  \ref{supfig:1}b. The modulated MW amplitude exhibits a near-Gaussian dependence on the IF voltage, likely due to the mixer nonlinearity.  In the experiment, the triangular wave has a duration of 1 \(\mu\)s, an amplitude of 200 mV, and a minimum voltage of 0 mV, producing a near-Gaussian pulse with an FWHM of 300 ns. A square wave is also applied to confine the RF output within the measurement window. When a triangular pulse with a peak voltage of 200 mV is fed into the IF port, the output MW pulse is shaped into a near-Gaussian wavepacket. Furthermore, since the mixer output is linearly proportional to the LO amplitude, the Gaussian pulse intensity can be controlled by adjusting the LO microwave power.

{ Next, we discuss the calibration of MW intensity and photon number in detail. Since the MOT lies in the far-field region of the antenna, the MW fields can be treated as plane waves relative to the small atomic medium. We first measure a moderate MW amplitude, $|E|=\hbar \Omega_{\mathrm{M}} /|\boldsymbol{d}_{34}|$, with an uncertainty of less than 1\% in the Autler-Townes splitting (ATS) regime\cite{Liao2020}. Here the MW amplitude corresponds to the intensity, $I_{\mathrm{M}}=\frac{|E|^{2}}{2}\sqrt{\frac{\epsilon_{0}}{\mu_{0}}}$, exceeding 4 pW mm$^{-2}$, where $\mu_{0}$ is the vacuum permeability. 
Using the field amplitudes, we calibrate the power meter reading from the MW generator and then extrapolate the low power setting for the intensities $I_{M} <$ 4 pW mm$^{-2}$. As mentioned in the data sheet of the signal generator (R\&S SMF100A), the output power nonlinearity is better than 1\% over the dynamic range. Taking into account the 1\% nonlinearity of the output power, the uncertainty of the low MW intensities extrapolated from the ATS regime would be less than 1.8\%.} By this method, the mean photon number of the input MW pulse is calibrated via the relation:
\begin{equation}
\bar N=\int I_M(t) d t A_\mathrm{L} /\hbar\omega_M,
\end{equation}
where $I_M(t)$ is the MW intensity with a Gaussian envelope in experiment. Once the LO amplitude is set, the input photon number is determined by using the Supplementary Equation (27). { Therefore, even at the single-photon level, we calibrate the average photon number of the  MW pulse with high precision.}

\section*{Supplementary Note 4 -- Other experimental parameter calibrations}

As in prior works\cite{Tu2022,JHan2018,Vogt2019, Borowka2024}, all auxiliary beams are focused onto the same end of the atomic ensemble. Due to the significantly larger beam waist of the blue light compared to the red light,  photon storage is confined to the red-light volume where all beams overlap. The MW incident angle with respect to the \(z\) axis is small, so the MW receiving area corresponds to the transverse region relative to the $z$-axis. The ensemble length is comparable to the Rayleigh range of the red-light beam. Accounting to the Gaussian atomic density distribution and the red beam's expansion along $z$, we derive the averaged MW storage cross-section below.

The longitudinal atomic density distribution is modeled as:
\begin{equation}
\tilde{n}(z)=n_{\mathrm{max}} e^{-2\left[(z-L/2)/ \textit{w}_{\text{A}}\right]^{2}} \theta\left(z\right) \theta\left(L-z\right),
\end{equation}
where $\theta\left(z\right)$ is the Heaviside step function, and $\textit{w}_{\text{A}}$ = $2L/3$ is the $1/e^2$ half width of the atomic density distribution. For a given optical depth $d_{\text{L}}$ and medium length $L$, the peak atomic density $n_{\mathrm{max}} $ is:
\begin{equation}
n_{\mathrm{max}}=\frac{{d_{\text{L}}} \Gamma_2}{ \bar{\beta}}\left(\int_{0}^{L} e^{-2\left[(z-L/2)/ \textit{w}_{\text{A}}\right]^{2}}  \mathrm{~d} z\right)^{-1},
\end{equation}
where $\bar{\beta}=2\omega_{\mathrm{P}}\left|\boldsymbol{d}_{21}\right|^{2} /\left( \hbar \epsilon_{0} c\right)$ and $\Gamma_2/2\pi$ = 6 MHz (natural linewidth of rubidium). The normalized density distribut is $\tilde{\rho}(z)$ = $\tilde{n}(z)/\mathcal{N}$, with the averaged density
\begin{equation}
\mathcal{N}=\frac{{d_{\text{L}}} \Gamma_2}{ \bar{\beta}L}.
\end{equation}

Despite beam expansion, the effective absorption volume is calculated by integrating cross-section along $z$. Since the coupling constant for field amplitude scales with $\propto \tilde{\rho}(z)$, the MW intensity absorption cross-section $I_{\mathrm{M}}$ scales with the square atomic density. Thus, the averaged cross-section is:
\begin{equation}
S_{\mathrm{M}}=\frac{1}{L} \int_{0}^{L}\pi r^{2}(z) \tilde{\rho}^2(z) \mathrm{~d} z,
\end{equation}
where the red beam radius at $z$ is:
\begin{equation}
r(z)=\textit{w}_{\mathrm{P}} \sqrt{1+\left(\frac{z}{R}\right)^{2}}.
\end{equation}
Here, $R$ is the Rayleigh range of the red-light beam. The mean beam radius is $\overline{\textit{w}}=\sqrt{S_{\mathrm{I}}/\pi}$. Using a beam profile (Thorlabs BC106N-VIS), we measure a red-light beam expansion from  54~$\mu$m (front) to 79~$\mu$m (rear) over $L=20$ mm, yielding a mean receiving-region radius of 66~$\mu$m.

To determine the medium OD, we use an acousto-optic modulator and scan the probe laser frequency through the $|1\rangle\rightarrow |2\rangle$ transition from -20 MHz to 20 MHz within 100 $\mu$s. When only a probe beam is incident on the medium, the global OD \(d_0\) is derived from fitting the two-level transmission spectrum,
\begin{equation}
I_{\text {out }}(\omega)=I_{\text {in }} \exp \left[-\operatorname{Im}\{\chi(\omega)\} k_P L\right],
\end{equation}
with the susceptibility given by
\begin{equation}
\chi(\omega)=-\frac{n_{at} \left|\mu_{12}\right|^2}{\varepsilon_0 \hbar\left(\omega+i \frac{\Gamma_2}{2}\right)}.
\end{equation}
The OD \(d_0\) is the only free parameter in the fitting, and its uncertainty is derived from 60 measurements. During the transduction window, the atoms are coherently trapped in the dressed state of a three-level system. {  For the Rydberg transition $|3\rangle \leftrightarrow |4\rangle$, we determine the MW OD by
\begin{equation}\label{d_M}
d_{\text{M}} = \frac{\left|\mu_{34}\right|^2 \lambda_P \Gamma_2\rho_{33}d_{0}}{\left|\mu_{12}\right|^2 \lambda_M \Gamma_4},
\end{equation}
where $\rho_{33}$ is the zero-order density element, $d_0$ is the OD of the transition \(|1\rangle \leftrightarrow |2\rangle \), and  $\Gamma_4\ (\Gamma_2)$ is the spontaneous decay rate of level $|4\rangle$ ($|2\rangle$). The typical parameters in our experiment are as follows: $|\mu_{34}| = 1271\, e a_0$, $\lambda_P = 780.2\ \mathrm{nm},$ $\Gamma_2 \approx 2\pi \times 6\ \mathrm{MHz},$ $\rho_{33} \approx \frac{|\Omega_P|^2}{|\Omega_P|^2 + |\Omega_A|^2} 
\approx 0.071$ (the Rabi frequencies \(\Omega_\mathrm{P}\), \(\Omega_\mathrm{A}\), and \(\Omega_\mathrm{R}\) are determined from their measured intensities), $d_0$ is measured to be $140 \pm 4$ , $|\mu_{12}| = 2.99\, e a_0,$ $\lambda_M \approx 8.0\ \mathrm{mm}$, $\Gamma_4 = 2\pi \times 740\ \mathrm{Hz}$. Therefore, from Supplementary Equation (\ref{d_M}), we obtain $d_\text{M}  \approx 7.5\times10^5.$

}

\section*{Supplementary Note 5 -- Rydberg dephasing induced by atom-atom interactions}

In the OMQT scheme, the EIT-based MW-memory is treated as a perturbation of the CPT system. Within the EIT-memory framework, polaritons composed of electromagnetic field and atomic coherence are decelerated are decelerated or accelerated by modulating auxiliary fields. Throughout this process, the atomic population remains in the dark state (a superposition of states $|3\rangle$ and $|5\rangle$), with minimal occupation state $|4\rangle$. In Supplementary Note 1, we have captured the photon transmission properties under the Rydberg dephasing (i.e. $\gamma_{51}$). Now in theory we quantify the dominated decoherence inherent to the dispersion of Rydberg energy level shifts\cite{Han2016}. 
For the cascade-type Rydberg system shown in Supplementary Figure 2a, the atoms interact with each other via the van der Waals (vdW) potentials $W_{33}=\hbar\left(C_6^{33}/ R^6\right) \hat{A}_{33}^1 \otimes \hat{A}_{33}^2$, $W_{55}=\hbar\left(C_6^{55} / R^6\right) \hat{A}_{55}^1 \otimes \hat{A}_{55}^2$, and $W_{35}=\hbar\left(C_6^{35} / R^6\right) \hat{A}_{33}^1 \otimes \hat{A}_{55}^2$, where $R$ is the interatomic separation, $\hat{A}_{\alpha \beta}^j \equiv\left|\alpha_j\right\rangle\left\langle\beta_j\right|$ ($\alpha,\beta \in {3,5}$) denote the atomic operators of atom $j=1,2$ and $C_6^{\alpha \beta}$ are the corresponding pair state vdW coefficients, which is calculated with the ARC software\cite{SibalicCPC2017}.

\begin{figure*}
\begin{center}
\includegraphics[width=14cm]{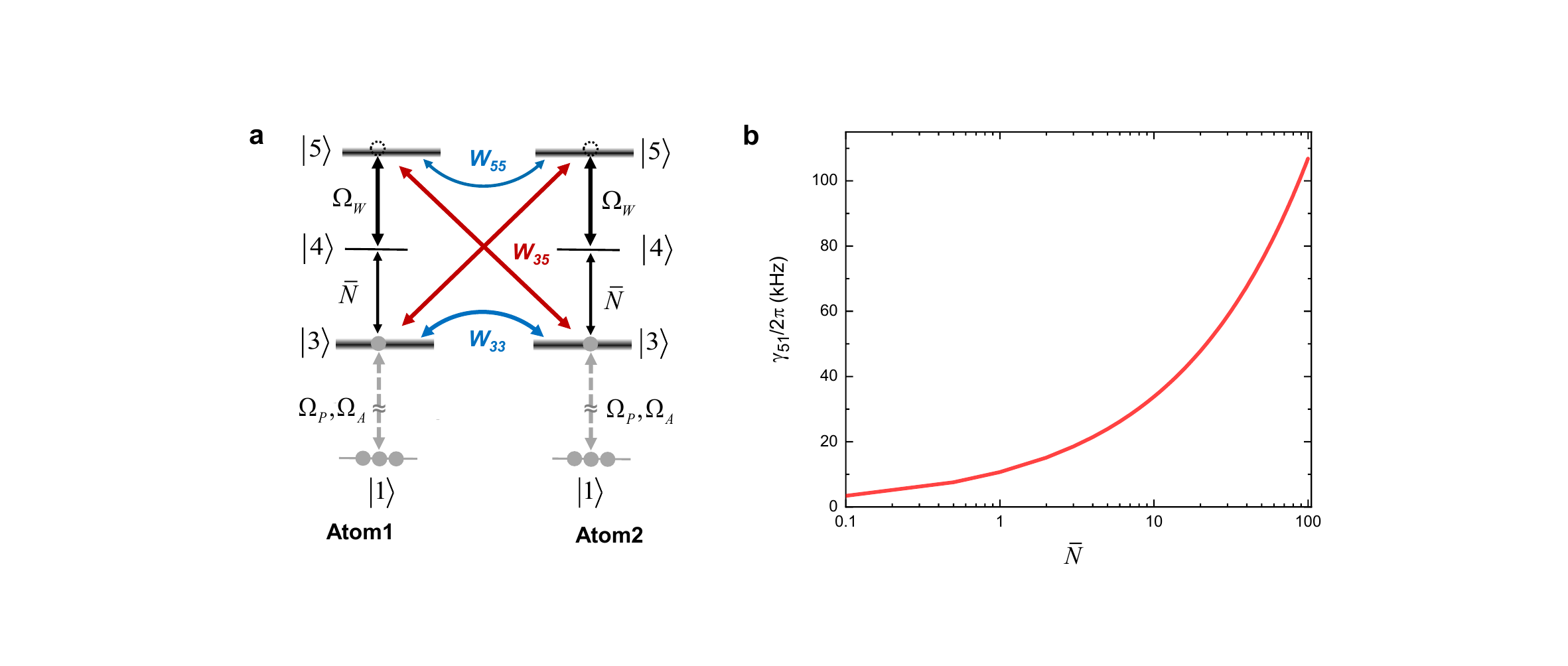}
\caption{\label{supfig:2}{\textbf{Rydberg dephasing in an interacting ensemble.} \textbf{a}, A coherent-population-trapping medium of strongly interacting Rydberg atoms coupled to an input MW with averaged photon number $\bar N$ and an auxiliary field with Rabi frequency $\Omega_{\rm{W}}$. The atoms in medium interact with each other via the $W_{33}$, $W_{35}$, and $W_{55}$ vdW interactions, where the subscript denotes the involved Rydberg pair-state. The relevant states are $|3\rangle=47S_{1/2}$, $|4\rangle=47P_{3/2}$, and $|5\rangle=46D_{5/2}$. \textbf{b}, Rydberg dephasing $\gamma_{51}$ as a function of the photon number $\bar N$. This figure depicts the theoretical results under the same condition as Figure 3a of the main text: $d\rm _{M}=0.75\times 10^6$, $\rho_{11} = 0.128$. When $\bar N$ = 1, we get the single-photon dephasing $\gamma_{0}/2\pi=10.8 \rm kHz$.}
}
\end{center}
\end{figure*}

By using the mean-field approach, we can simulate the Rydberg dephasing with the standard deviation of energy level shift resulted from the vdW interactions\cite{Han2016}. To approximately merge the spatial correlations between Rydberg atoms, we take a short-range cutoff to spatial integral at the blockade radius $R_B$. For the $47S_{1/2}47S_{1/2}$ pair-state, the $W_{33}$ interaction leads to a bandwidth of the $|3\rangle$ level shift, which is defined as $2\sqrt{\vartheta _{33}}$, with a level shift variance
\begin{equation}
\begin{aligned}
\vartheta _{33}&=  \overline{(\sum_j \frac{C_6^{33}}{R_{i j}^6})^{2}} -V_{\mathrm{vdW}}^{2}\\
&\approx \int_{R_B}^{\infty} \rho_{33} n_{at} (\frac{C_6^{33}}{r^{6}})^2 4 \pi r^2 d r\\
&= \frac{4 \pi (C_6^{33})^2  \rho_{33} n_{a t}}{9 R_B^9}.
\end{aligned}
\end{equation}
Here $R_B=\left(2 C_6^{33} / \gamma_{\mathrm{pump}}\right)^{1 / 6}=4.1 \mu \mathrm{m}$, $\gamma_{\mathrm{pump}}/2/\pi =3 \rm MHz$ is the measured linewidth of the pumping Rydberg EIT, and $n_{at}=2.4 \times 10^{10} \mathrm{~cm}^{-3}$ is the total atomic number density.

According to Supplementary Equation (25), only the dephasing term associated with level $|5\rangle$ substantively degrades the transmission efficiency. When few of MW photons are stored in the medium, there are negligible atoms occupied in the state $|5\rangle$. Therefore, the $46D_{5/2}46D_{5/2}$ pair-state possesses an ignorable energy level shift and bandwidth due to the $W_{55}$ interaction. Furthermore, the population of state $\left|5 \right\rangle$ is approximately equal to the averaged photon number $\bar N$ of input MW in the high-efficiency storage. In this case, the population of state $\left|3 \right\rangle$ is unchanged during the few-photon transduction, and the $47S_{1/2}46D_{5/2}$ pair-state constrains the medium to behave as the long-range interactions between the $\bar N$ excitations and the Rydberg-state populations of $47S_{1/2}$. Consequently, the $W_{35}$ interaction dominates the Rydberg dephasing when we quantify the transmission property in storage process, and then
\begin{equation}
\begin{aligned}
\gamma_{51}&\approx \sqrt{\vartheta _{55}}+\sqrt{\vartheta _{35}} \\
&\simeq \sqrt{\frac{4 \pi (\bar{C}_6^{35})^2  \bar N\rho_{33} n_{a t}}{9 R_B^9}}\\
&= \sqrt{\bar{N}}\gamma_{0},
\end{aligned}
\end{equation}
where $\bar{C}_6^{35}$ = 0.15 GHz $\mu$m$^6$ is the mean vdW coefficient obtained by integrating the angular distribution of $C_6^{35}$, and we have defined the single-photon dephasing as
\begin{equation}
\gamma_{0}= \sqrt{\frac{4 \pi (\bar{C}_6^{35})^2  \rho_{33} n_{a t}}{9 R_B^9}}.
\end{equation}

Then we discuss the Rydberg dephasing induced by dipole-dipole exchange (DDE) interactions between the atoms. The MW fields drive resonant Rydberg transitions during EIT-storage, leading to two types of DDE interactions involving  level $|5\rangle$, i.e. $D_{45}=\hbar\left(C_3/ R^3\right) \hat{A}_{45}^1 \otimes \hat{A}_{54}^2$ and $D_{(34,45)}=\hbar\left(C^{\prime}_3/ R^3\right) \hat{A}_{34}^1 \otimes \hat{A}_{45}^2$. For the $D_{45}$ interaction, the Rydberg energy level shifts due to DDE coupling, with an eigenvalue $E_{45}$. The variance of this DDE interaction is given by
\begin{equation}
\begin{aligned}
\vartheta _{45}&=  \overline{(\sum_j \frac{C_3}{R_{i j}^3})^{2}} -E_{45}^{2}\\
&\approx \int_{R_B}^{w}    \frac{|P_{41}P^{*}_{51}|}{\rho_{11}}\frac{(C_3)^{2}}{r^{6}} n_{a t}4 \pi r^2  d r\\
&\simeq {{4\pi (C_3)^{2}{|P_{41}P^{*}_{51}|}{n_{at}}} \over {3\rho_{11}R_B^3}},
\end{aligned}
\end{equation}
where the $C_3$ coefficient is about 0.29 GHz $\mu$m$^3$, and $w$ is the $1/e^{2}$ radius of coupling laser beam. Since the input MW field is extremely weak ($|\Omega _M| \ll |\Omega _W|$), $|P_{41}P^{*}_{51}| \ll \rho_{33}$ and thus $\vartheta _{45} \ll \vartheta _{35}$ for few-photon storage. The DDE interaction of $D_{(34,45)}$ only occurs during the  two-photon transition from state $|3\rangle$ to $|5\rangle$, where the write field $\Omega _W$ ramps down to zero within about 10 ns. The ramp-down period is much shorter than the group delay time $t_d$, so the accumulated decoherence  from $D_{(34,45)}$ is negligible compared to $D_{45}$ coupling. Therefore, we disregard Rydberg dephasing terms induced by the DDE interactions.

Note that in conventional memory schemes based on a $\Lambda$-type system, off-resonant excitations of the adjacent excited state can introduce multiphoton loss channels or multi-wave mixing \cite{Hsiao2018}, which degrades the storage efficiency. However, Our memory approach is immune to this adverse effect for two key reasons. First, the sign-matched circularly-polarized fields couple the atomic transitions to the maximal magnetic sublevels. So our memory scheme utilizes a set of closed transitions during the write process, thereby preventing cross-talk between nearby Rydberg states. Second, any off-resonant excitation from level $|3\rangle$ by the write pulse is negligible, as the frequency detuning exceeds ten GHz, which prevents degradation in memory fidelity.  In general, the overall dephasing should take an account of finite laser linewidth, stray fields, blackbody radiation, and Rydberg state lifetime (about tens of $\mu s$), but those terms are much smaller than the above Rydberg dephasing rate\cite{Tu2023}. The Rydberg dephasing $\gamma_{51}$ as a function of the photon number $\bar N$ is show in Supplementary Figure 2b. When $\bar N$ = 1, we get the single-photon dephasing $\gamma_{0}/2\pi=10.8 \rm kHz$.

\section*{Supplementary Note 6 -- Thermal background radiation}
The OMQT can store the ambient thermal MW photons and converts them into the optical photons. Unlike the input MW signal, blackbody radiation is unpolarized and isotropic. For a given mode, the mean photon number is given by
\begin{equation}
\bar n (\nu, T)= \frac{1}{e^{h \nu / k T}-1},
\end{equation}
yielding approximately 170 photons at temperature $T\simeq$ 300 K and frequency $\nu$ of 37 GHz. Using the Stefan-Boltzmann law, the radiation photon flux incident on  the medium surface $A$ is
\begin{equation}
\Phi(\nu, T)=\int d \nu \int_{\Omega_{0}} d \Omega \frac{2 \nu^{2}\bar n (\nu, T)}{c^{2}} \cos \theta \cdot A,
\end{equation}
where $d \Omega=\sin \theta d \theta d \varphi$ is the unit solid angle, and $A=2 \pi r^{2}+2 \pi r L$ is the surface area of the cylindric interaction medium (radius  $r=66 \mu$m and length $L$ = 20 mm). Integrating over the 2.1 MHz storage bandwidth, the thermal photon flux is $\Phi\approx 295$ MHz. Since only $\sigma^{+}$ circularly polarized MW photons are converted in our scheme, the number of thermal photons participating in the storage process within the interaction time $\tau \simeq$ 550 ns is:
\begin{equation}
N=\frac{\Phi \times \tau}{2} \approx 81.2.
\end{equation}

\begin{figure*}[t]
\begin{center}
\includegraphics[width=17cm]{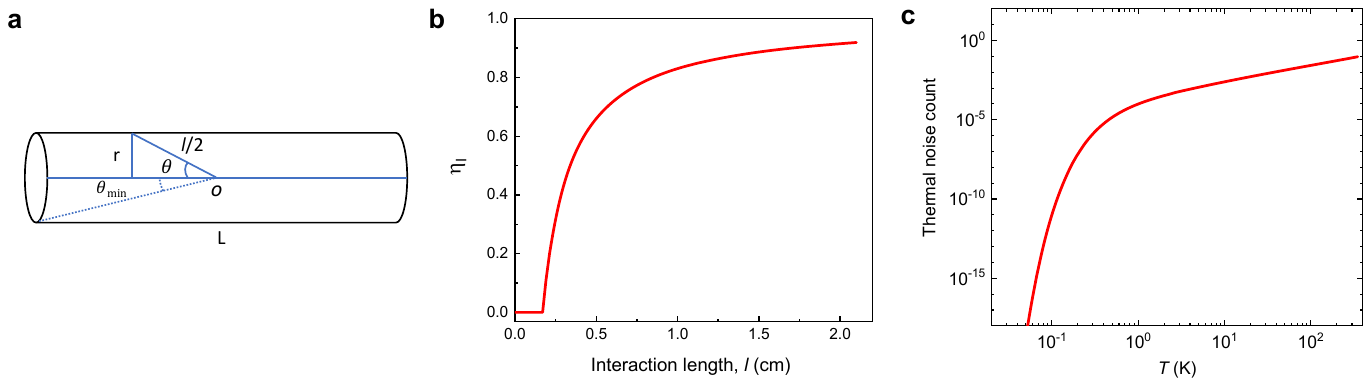}
\caption{\label{BBR} \textbf{Photon noise attributed to the thermal MW background.} \textbf{a}, The model of a cylindric interaction volume. \textbf{b}, Storage efficiency $\eta_\mathrm{I}$ versus the interaction length $l$. \textbf{c}, Thermal noise count $\bar n_{th}$ versus the ambient temperature $T$.
 }
\end{center}
\end{figure*}

Although many thermal photons are involved, only a small fraction propagating within an extremely narrow solid angle relative to $z$-axis are efficiently stored in the cigar-shaped medium ($r \ll L$). As illustrated in Supplementary Figure 3a, we set the origin $O$ as the medium center and calculate the thermal photon contribution  from angle $\theta$ as $d S= \frac{N \sin \theta d \theta}{2}$.  The effective interaction length $l$ for thermal photons at angle $\theta$ is
\begin{equation}
l=\frac{2 r}{\sin \theta}, \quad \theta \epsilon\left[\theta_{\min }, \pi-\theta_{\min }\right],
\end{equation}
where $\theta_{\min }=\arcsin \left(\frac{2 r}{L}\right)$   $\sim 10^{-4}$ , and, $l \simeq L$ just for the paraxial thermal photons radiated on the end surface of interaction volume.

The storage efficiency $\eta_\mathrm{I}$ decreases sharply as the interaction length shortens. The dependence of $\eta_\mathrm{I}$ on the length $l$ is calculated and shown Supplementary Figure 3b. For our cigar-shaped medium, only thermal photons within a solid angle of 0.077 radians achieve efficiency $\eta_\mathrm{I}>$ 0.5 $\%$. Taking into account the SE, the number of optical photons converted from  MW background radiation is:
\begin{equation}
\label{BBREQ}
\begin{aligned}
\bar n_{th} &=\int_{\theta_{\min }}^{\pi-\theta_{\min }} \eta_\mathrm{I}(l) d S+\frac{2 \eta_{\max}N \int_{0}^{\theta \min } \sin \theta d \theta}{4 \pi} \\
&=\int_{\theta_{\min }}^{\pi-\theta_{\min }} \frac{\eta_\mathrm{I}(\theta) N \sin \theta d \theta}{2}+\frac{2 \eta_{\max } N \int_{0}^{\theta_{\min}} \sin \theta d \theta}{4 \pi} ,
\end{aligned}
\end{equation}
where $\eta_{\max}$ is the peak $\eta_\mathrm{I}$ and can be considered equivalent to $\eta_0$ in Eq.(2) in the main text. We find that the thermal noise count $\bar n_{th} \approx 0.08 $ (0.109) for $\eta_{\max}=91\%\ (93\%) $. This theoretical result agrees with the observed thermal noise count (0.109 per pulse). Supplementary Figure 3c shows the thermal photon count as a function of temperature $T$, demonstrating that the thermal noise count falls below $10^{-4}$ per pulse in a millikelvin cryogenic environment. In short, although there are approximately 170 thermal photons at 37 GHz and 300 K, our theoretical calculations show that only approximate  0.1 photons are converted into optical phonons by the OMQT. This is because only the paraxial background radiation propagating within a tiny solid angle can be efficiently stored in our cigar-shaped medium and converted into optical photons.

\begin{figure}[t]
\begin{center}
\includegraphics[width=12cm]{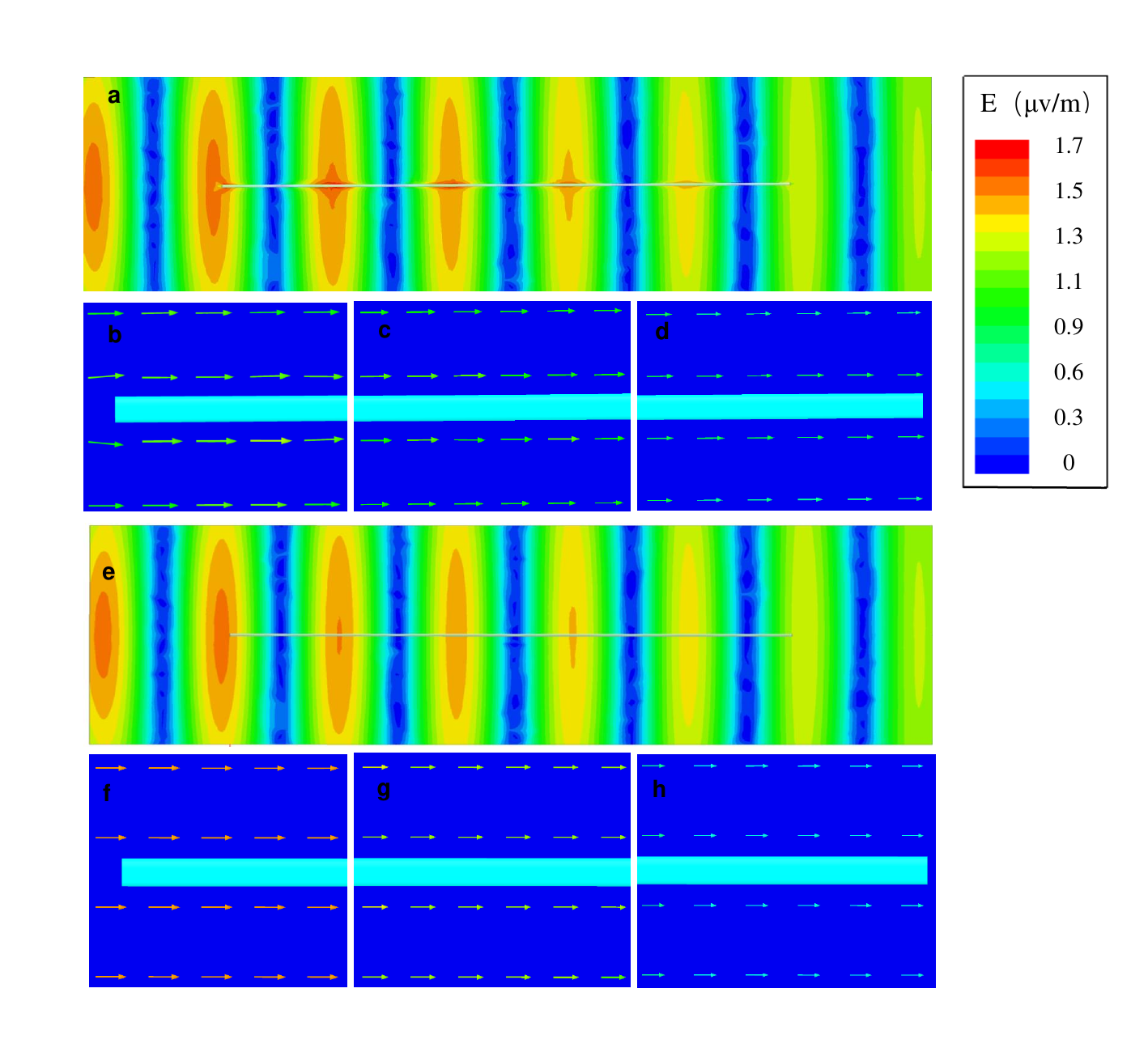}
\caption{\label{supfig:4}{ 
\textbf{Simulation of the MW electric field treated as a plane wave incident upon a cigar-like subwavelength medium.}
\textbf{a}, Global view for a dielectric material with an effective permittivity and loss tangent similar to OMQT.
\textbf{b--d}, Partial views of the front-end, middle-end, and back-end regions in \textbf{a}.
\textbf{e}, Global view for a dielectric material with the same relative permittivity and loss tangent as air.
\textbf{f--h}, Partial views of the front-end, middle-end, and back-end regions in \textbf{e}.
The medium is the same size as OMQT, and the effective permittivity and loss tangent in \textbf{a} are derived from OMQT parameters, such as microwave delay time and optical depth. The arrow represents the Poynting vector of the electromagnetic wave.
  }}
\end{center}
\end{figure}

{ Next, we assess the validity of our geometric approach for modeling MW absorption by a subwavelength atomic medium. To precisely determine the flux-density, the atomic ensemble used in our experiment resides in the far-field region, and thus the MW field to be converted can be approximately regarded as a plane wave. However, the interaction region is deeply subwavelength for the MW field in the transverse dimension, yet comparable to the MW wavelength in the longitudinal direction. In order to verify the uniformity of flux density and assess the impact of MW diffraction, we simulate the electromagnetic waves near the cigar-like medium using the software package \textit{High Frequency Structure Simulator}. Since storing the MW pulse is a dynamic process, we derive the effective permittivity and loss tangent from OMQT parameters, such as MW delay time and optical depth. Supplementary Figure 4 shows that despite strong absorption, the MW field near the medium maintains good paraxial conditions. Compared to the air case (see Supplementary Figure 4e-h), the diffraction effect of the atomic medium is negligible. Unlike a high-gain, highly directional classical horn antenna, the dielectric only slightly perturbs the electromagnetic wave. This argument also matches well with our experimental measurements of the MW intensity based on the Rydberg EIT-ATS method, confirming the stability of the MW flux density in the far-field region.
Despite such high efficiency within a tiny receiving angle, the MW near the cigar-like medium exhibits negligible diffraction effects, and still behaves as a plane wave with a highly uniform and stable electric field. Therefore, our geometric treatments for thermal radiation and the MW field are reasonable.}

{ Finally, we discuss the influence of phase mismatching during the storage process. When the write field is turned off, the MW photons are mapped as the spin wave of atomic ensemble, yielding a spin-wave vector $\mathbf{k}_{\mathrm{SW}}=\mathbf{k}_{\mathrm{A}}-\mathbf{k}_{\mathrm{P}}+\mathbf{k}_{\mathrm{M}}-\mathbf{k}_{\mathrm{W}}$. As mentioned in the Methods, this residual momentum mainly degrades the spin-wave coherence, thus reducing the storage time, but does not affect the storage efficiency. For short-duration storage, the efficiency primarily depends on the optical thickness experienced by the MW along its propagation path. For photons incident at an angle greater than $\theta_{\min}$, the short interaction length sharply decreases the storage efficiency, as shown in Supplementary Equation (\ref{BBREQ}). Moreover, the phase-matching condition is substantially insensitive to the MW incident angle due to the vast disparity between MW and optical wavevectors, thereby imposing no additional constraints on the receiving angle beyond geometric overlap. In our system, $k_{\mathrm{sw}} \approx 5 \times 10^6 \mathrm{~m}^{-1}$. When the MW field direction reverses from $0^{\circ}$ to $180^{\circ}$, altering the spin-wave vector by $2\mathbf{k}_{\mathrm{M}}$, the maximum variation (i.e., $2 \mathbf{k}_{\mathrm{M}} \approx 1.6 \times 10^3 \mathrm{~m}^{-1}$ for 37.5 GHz) amounts to only approximately 0.03\% of the spin-wave vector. Therefore, the emission angle deviation is well within the collection NA of standard optics and contributes negligibly to the efficiency loss. }

\section*{Supplementary Note 7 -- Remarks on the definition of area-normalized efficiency }

We here demonstrate that the area-normalized efficiency defined in Supplementary Equation (26) is equivalent to the flux-density method under the condition \(A_L = A_R\), where \(A_R\) represents the effective beam area for photon retrieval.  This condition is inherently  satisfied  in the collinear configuration employed in our study.  For a pulse duration  \(t_0\), the intensity of the converted field is:
\begin{equation}
{I_L} = \frac{N_L {\hbar \omega_L}}{A_R t_0},
\end{equation}
where $N_L$ is the number of converted photons. The mean MW intensity is:
\begin{equation}
\bar{I}_M=\frac{\int I_M(t) \, dt}{t_0}.
\end{equation}
Therefore, we find an intensity-to-intensity efficiency $\eta_{ii}$
\begin{equation}
\eta_{ii} = \frac{I_L / \hbar \omega_L}{\bar{I}_M / \hbar \omega_M} = \frac{N_L \hbar \omega_M}{A_R \int I_M(t) \, dt}=\eta_\mathrm{I},
\label{eq:efficiency}
\end{equation}
under the condition \(A_R = A_L\). In this case, we establish the linear relationship between MW intensity (\(I_M\)) and source power, along with the averaged cross-section (\(A_L\)). By substituting \(A_L = A_R\) into Supplementary Equation (27), we precisely tune the average MW photon number \(\bar{N}\) from 0.1 to 130 by adjusting the MW source power.

It demonstrated that the area-normalized efficiency remains a practical physical metric for characterizing the conversion efficiency in free-space configurations, although the interaction region is substantially subwavelength for the MW fields. As experimentally demonstrated in Ref.\cite{Borowka2024}, the measured $\eta_\mathrm{I}$ in their system agrees with theoretical expectations. Similarly, in our experimental setup, the measured $\eta_\mathrm{I}$ aligns closely with the theoretical derivation detailed in Supplementary Note 1.

{  Finally, we discuss the mode-matching losses in free-space transduction. Since the medium is in the far-field region of the antenna horn (as shown in Fig. 2b of the main text), the MW power is coupled from the antenna to the atomic ensemble with low efficiency, and this part dominates the total end-to-end efficiency of our current setup. The atomic ensemble is at a distance of 
$d$=30 cm from the antenna horn, which has a gain $g=20$dB. For a beam radius of 
$r$ = 66 $\mu$m, the MW photons emitted from the horn are coupled into the medium with an efficiency of $\frac{g \pi r^{2}}{4\pi d^{2}}\simeq 1.2 \times 10^{-6}.$ In addition, for the OMQT and readout path, the entire chain yields an end-to-end efficiency of 4.3 $\times$ $10^{-7}$  from the antenna horn to the single-photon detector.


Current experimental demonstrations have already shown devices that couple Rydberg ensembles to confined MW modes or trap atoms in cryogenic environments~\cite{FortaghNC2017, FortaghPRapp2025, Kumar2023}. By leveraging these technological advancements, our setup can be upgraded to operate in a cryogenic environment. In the waveguide integration scheme, MW photons propagate in a single mode along the waveguide, above which the ensemble is trapped. This updated setup mitigates mode-matching losses, thereby significantly enhancing the overall end-to-end efficiency. Specifically, a cigar-shaped ensemble of cold atoms can be trapped in an optical trap~\cite{FortaghPRapp2025} or a magnetic trap using a Z-shaped wire~\cite{FortaghNC2017}. Because the ensemble has a large optical thickness, an optical cavity is not necessarily required in our scheme. This cavity-free design makes the experimental setup less sensitive to vibrations from the dilution refrigerator, thus rendering it more compatible with a cryogenic environment. }

\endgroup   


\begin{thebibliography}{99}

\bibitem{KimbleNature2008} Kimble, H. J. The quantum internet. \textit{Nature} \textbf{453}, 1023-1030 (2008).
\bibitem{WehnerSci2018} Wehner, S., Elkouss, D., Hanson, R. Quantum internet: A vision for the road ahead. \textit{Science} \textbf{362}, eaam9288 (2018).
\bibitem{ClarkNature2008} Clarke, J., Wilhelm, F. K. Superconducting quantum bits. \textit{Nature} \textbf{453}, 1031 (2008).
\bibitem{DevoretScience2013} Devoret, M. H., Schoelkopf, R. J. Superconducting circuits for quantum information: an outlook. \textit{Science} \textbf{339}, 1169 (2013).
\bibitem{PompiliScience2021} Pompili, M. et al. Realization of a multinode quantum network of remote solid-state qubits. \textit{Science} \textbf{372}, 259-264 (2021).
\bibitem{ArquerScience2021} Arquer, F. P. G. D. et al. Semiconductor quantum dots: Technological progress and future challenges. \textit{Science} \textbf{373}, eaaz8541 (2021).
\bibitem{LambertReview2019} Lambert, N.J., Rueda, A., Sedlmeir, F., Schwefel, H.G.L. Coherent conversion between microwave and optical photons-an overview of physical implementations. \textit{Adv. Quantum Technol.} \textbf{3}, 1900077 (2020).
\bibitem{LaukReview2020} Lauk, N. et al. Perspectives on quantum transduction. \textit{Quantum Sci. Technol.} \textbf{5}, 020501 (2020).
\bibitem{XHan2021} Han, X., Fu, W., Zou, C. L., Jiang, L., Tang, H. Microwave-optical quantum frequency conversion. \textit{Optica} \textbf{8}, 1050 (2021).
\bibitem{XiangRMP2013} Xiang, Z. L. et al. Hybrid quantum circuits: Superconducting circuits interacting with other quantum systems. \textit{Rev. Mod. Phys.} \textbf{85}, 623 (2013).
\bibitem{Duan2001} Duan, L. M., Lukin, M. D., Cirac, J. I., Zoller, P. Long-distance quantum communication with atomic ensembles and linear optics. \textit{Nature} \textbf{414}, 413 (2001).
\bibitem{Sangouard2011} Sangouard, N., Simon, C., DeRiedmatten, H., Gisin, N. Quantum repeaters based on atomic ensembles and linear optics. \textit{Rev. Mod. Phys.} \textbf{83}, 33 (2011).
\bibitem{LvovskyNP2009} Lvovsky, A., Sanders, B., Tittel, W. Optical quantum memory. \textit{Nat. Photon.} \textbf{3}, 706--714 (2009).
\bibitem{BochmannNP2013} Bochmann, J., Vainsencher, A., Awschalom, D. D., Cleland, A. N. Nanomechanical coupling between microwave and optical photons. \textit{Nat. Phys.} \textbf{9}, 712-716 (2013).
\bibitem{HigginbothamNP2018} Higginbotham, A. et al. Harnessing electro-optic correlations in an efficient mechanical converter. \textit{Nat. Phys.} \textbf{14}, 1038-1042 (2018).
\bibitem{ForschNP2019} Forsch, M., Stockill, R., Wallucks, A., Marinkovi, I., Grblacher, S. Microwave-to-optics conversion using a mechanical oscillator in its quantum ground state. \textit{Nat. Phys.} \textbf{16}, 69-74 (2019).
\bibitem{Han2018} Han, J. et al. Coherent microwave-to-optical conversion via six-wave mixing in Rydberg atoms. \textit{Phys. Rev. Lett.} \textbf{120}, 093201 (2018).
\bibitem{Vogt2019} Vogt, T. et al. Efficient microwave-to-optical conversion using Rydberg atoms. \textit{Phys. Rev. A} \textbf{99}, 023832 (2019).
\bibitem{Tu2022} Tu, HT. et al. High-efficiency coherent microwave-to-optics conversion via off-resonant scattering. \textit{Nat. Photon.} \textbf{16}, 291 (2022).
\bibitem{Kumar2023} Kumar, A. et al. Quantum-enabled millimetre wave to optical transduction using neutral atoms. \textit{Nature} \textbf{615}, 614 (2023).
\bibitem{Borowka2024} Borowka, S., Pylypenko, U., Mazelanik, M. P. M. Continuous wideband microwave-to-optical converter based on room-temperature Rydberg atoms. \textit{Nat. Photon.} \textbf{18}, 32--38 (2024).
\bibitem{RuedaOptica2016} Rueda, A. et al. Efficient microwave to optical photon conversion: an electro-optical realization. \textit{Optica} \textbf{3}, 597-604 (2016).
\bibitem{FanSA2018} Fan, L., Zou, C. L., Cheng, R., Guo, X., Han, X. Superconducting cavity electro-optics: A platform for coherent photon conversion between superconducting and photonic circuits. \textit{Sci. Adv.} \textbf{4}, eaar4994 (2018).
\bibitem{WilliamsonPRL2014} Williamson, L. A., Chen, Y. H., Longdell, J. J. Magneto-optic modulator with unit quantum efficiency. \textit{Phys. Rev. Lett.} \textbf{113}, 203601 (2014).
\bibitem{Kurokawa2022} Kurokawa, H., Yamamoto, M., Sekiguchi, Y., Kosaka, H. Remote Entanglement of Superconducting Qubits via Solid-State Spin Quantum Memories. \textit{Phys. Rev. Applied} \textbf{18}, 064039 (2022).
\bibitem{Saito2013} Saito, S., Zhu, X., Robert Amsuss, Matsuzaki, Y., Semba, K. Towards realizing a quantum memory for a superconducting qubit: storage and retrieval of quantum states. \textit{Phys. Rev. Lett.} \textbf{111}, 107008 (2013).
\bibitem{Julsgaard2013} Julsgaard, B., Grezes, C., Bertet, P., Molmer, K. Quantum memory for microwave photons in an inhomogeneously broadened spin ensemble. \textit{Phys. Rev. Lett.} \textbf{110}, 250503 (2013).
\bibitem{Ranjan2020} Ranjan, V. et al. Multimode storage of quantum microwave fields in electron spins over 100 ms. \textit{Phys. Rev. Lett.} \textbf{125}, 210505 (2020).
\bibitem{Flurin2015} Flurin. E. et al. Superconducting quantum node for entanglement and storage of microwave radiation. \textit{Phys. Rev. Lett.} \textbf{114}, 090503 (2015).
\bibitem{Reagor2016} Reagor. M. et al. Quantum memory with millisecond coherence in circuit QED. \textit{Phys. Rev. B} \textbf{94}, 014506 (2016).
\bibitem{Pfaff2017} Pfaff, W. et al. Controlled release of multiphoton quantum states from a microwave cavity memory. \textit{Nat. Phys.} \textbf{13}, 882 (2017).
\bibitem{Bao2021} Bao, Z. et al. On-demand storage and retrieval of microwave photons using a superconducting multiresonator quantum memory. \textit{Phys. Rev. Lett.} \textbf{127}, 010503 (2021).
\bibitem{Palomaki2013} Palomaki, T. A., Harlow, J. W., Teufel, J. D., Simmonds, R. W., Lehnert, K. W. Coherent state transfer between itinerant microwave fields and a mechanical oscillator. \textit{Nature} \textbf{495}, 210 (2013).
\bibitem{Liu2023} Liu, Y. et al. Coherent memory for microwave photons based on long-lived mechanical excitations. \textit{npj Quantum Inf.} \textbf{9}, 80 (2023).
\bibitem{Hsiao2018} Hsiao, Y.-F. et al. Highly efficient coherent optical memory based on electromagnetically induced transparency. \textit{Phys. Rev. Lett.} \textbf{120}, 183602 (2018).
\bibitem{Vernaz2018} Vernaz-Gris, P. et al. Highly-efficient quantum memory for polarization qubits in a spatially-multiplexed cold atomic ensemble. \textit{Nat. Commun.} \textbf{9}, 363 (2018).
\bibitem{Distante2017} Distante, E. et al. Storing single photons emitted by a quantum memory on a highly excited Rydberg state. \textit{Nat. Commun.} \textbf{8}, 14072 (2017).
\bibitem{Wang2019} Wang, Y. et al. Efficient quantum memory for single-photon polarization qubits. \textit{Nat. Photon.} \textbf{13}, 346 (2019).
\bibitem{KYSu2022} Su, K.Y. et al. Quantum Interference between Nonidentical Single Particles. \textit{Phys. Rev. Lett.} \textbf{129}, 093604 (2022).
\bibitem{KiffnerNJP2016} Kiffner, M. et al. Two-way interconversion of millimeter-wave and optical fields in Rydberg gases. \textit{New J. Phys.} \textbf{18}, 093030 (2016).
\bibitem{FortaghPRR2022} Kaiser, M. et al. Cavity driven Rabi oscillations between Rydberg states of atoms trapped on a superconducting atom chip. \textit{Phys. Rev. Research} \textbf{4}, 013207 (2022).
\bibitem{FortaghNC2017} Hattermann, H. et al. Coupling ultracold atoms to a superconducting coplanar waveguide resonator. \textit{Nat. Commun.} \textbf{8}, 2254 (2017).
\bibitem{FortaghPRapp2025} Wilde, B. et al. Superconducting on-chip microwave cavity for tunable hybrid systems with optically trapped Rydberg atoms. \textit{Phys. Rev. Applied} \textbf{23}, 064016 (2025).
\bibitem{RMQM1} Reim, K., Nunn, J., Lorenz, V. et al. Towards high-speed optical quantum memories. \textit{Nature Photon.} \textbf{4}, 218-221 (2010).
\bibitem{RMQM2} Guo, J., Feng, X., Yang, P. et al. High-performance Raman quantum memory with optimal control in room temperature atoms. \textit{Nat. Commun.} \textbf{10}, 148 (2019).
\bibitem{Gorshkov2007a} Gorshkov, A. V., Andre, A., Fleischhauer, M., Sorensen, A. S., Lukin, M. D. Universal Approach to optimal photon storage in atomic media. \textit{Phys. Rev. Lett.} \textbf{98}, 123601 (2007).
\bibitem{Li2016} Li, L., Kuzmich, A. Quantum memory with strong and controllable Rydberg-level interactions. \textit{Nat. Commun.} \textbf{7}, 13618 (2016).
\bibitem{Axline2018} Axline, C.J. et al. On-demand quantum state transfer and entanglement between remote microwave cavity memories. \textit{Nature Phys.} \textbf{14}, 705--710 (2018).
\bibitem{LandraPRA2019} Landra, A. et al. Design of an experimental platform for hybridization of atomic and superconducting quantum systems. \textit{Phys. Rev. A} \textbf{99}, 053421 (2019).
\bibitem{Komatsu2022} Komatsu, E. New physics from the polarized light of the cosmic microwave background. \textit{Nat. Rev. Phys.} \textbf{4}, 452-469 (2022).
\bibitem{Schlossberger2024} Schlossberger, N., Prajapati, N., Berweger, S. et al. Rydberg states of alkali atoms in atomic vapour as SI-traceable field probes and communications receivers. \textit{Nat. Rev. Phys.} \textbf{6}, 606-620 (2024).
\bibitem{KurpiersPRApp2019} Kurpiers, P., et al. Quantum Communication with Time-Bin Encoded Microwave Photons. \textit{Phys. Rev. Appl.} \textbf{12}, 044067 (2019).
\bibitem{Bryan2017} Gard, B. T., Jacobs, K., McDermott, R., Saffman, M. Microwave-to-optical frequency conversion using a cesium atom coupled to a superconducting resonator. \textit{Phys. Rev. A} \textbf{96}, 013833 (2017).
\bibitem{SibalicCPC2017} \v{S}ibali\'{c}, N., Pritchard, J. D., Adams, C. S., Weatherill, K. J. ARC: An open-source library for calculating properties of alkali Rydberg atoms. \textit{Comput. Phys. Comm.} \textbf{220}, 319-331 (2017).
\bibitem{Liao2020} Liao, K.Y. et al. Microwave electrometry via electromagnetically induced absorption in cold Rydberg atoms. \textit{Phys. Rev. A} \textbf{101}, 053432 (2020).
\bibitem{XHLiu2022} Liu, X. H. et al. Continuous-Frequency Microwave Heterodyne Detection in an Atomic Vapor Cell. \textit{Phys. Rev. Applied} \textbf{18}, 054003 (2022).
\bibitem{Tu2023} Tu, H. et al. Approaching the standard quantum limit of a Rydberg-atom microwave electrometer. \textit{Sci. Adv.} \textbf{10}, eads0683 (2024).

\end{thebibliography}

\begin{thebibliography}{99}

\bibitem{Hsiao2018} Hsiao.Y.-F. et al.  Highly efficient coherent optical memory based on electromagnetically induced transparency. Phys. Rev. Lett. \textbf{120}, 183602 (2018).
\bibitem{Borowka2024} Borowka, S.,  Parniak, M., Pylypenko, U., Mazelanik, M., Continuous wideband microwave-to-optical converter based on room-temperature Rydberg atoms.  Nat. Photon. \textbf{18}, 32 (2024).
\bibitem{Tu2022} Tu, H. et al. High-efficiency coherent microwave-to-optics conversion via off-resonant scattering.  \textit{Nat. Photon.} \textbf{16}, 291 (2022).
\bibitem{JHan2018}  Han, J. et al. Coherent microwave-to-optical conversion via six-wave mixing in Rydberg atoms. Phys. Rev. Lett. \textbf{120}, 093201 (2018).
\bibitem{Vogt2019} Vogt, T. et al. Efficient microwave-to-optical conversion using Rydberg atoms. Phys. Rev. A \textbf{99}, 023832 (2019).
\bibitem{Liao2020}  Liao, KY. et al. Microwave electrometry via electromagnetically induced absorption in cold Rydberg atoms. \textit{Phys. Rev. A} \textbf{101}, 053432 (2020).
\bibitem{XHLiu2022} X. H. Liu et al., Continuous-Frequency Microwave Heterodyne Detection in an Atomic Vapor Cell, \textit{Phys. Rev. Applied.} \textbf{18}, 054003 (2022).
{ \bibitem{Wang2019}Wang, Y. et al. Efficient quantum memory for single-photon polarization qubits. \textit{Nat. Photon.} \textbf{13}, 346 (2019).
\bibitem{SibalicCPC2017}  \v{S}ibali\'{c}, N., N., Pritchard, J. D., Adams, C. S.,  Weatherill, K. J.  ARC: An open-source library for calculating properties of alkali Rydberg atoms. \textit{Comput. Phys. Comm.} \textbf{220} 319-331 (2017).}
\bibitem{Han2016}  Han, J.,  Vogt. T., and Li, W., Spectral shift and dephasing of electromagnetically induced transparency in an interacting Rydberg gas. \textit{Phys. Rev. A} \textbf{94}, 043806 (2016).



\bibitem{Tu2023} Tu, H. et al. Approaching the standard quantum limit of a Rydberg-atom microwave electrometer. \textit{Sci. Adv.} \textbf{10}, eads0683 (2024).

{ \bibitem{FortaghNC2017} Hattermann, H. et al. Coupling ultracold atoms to a superconducting coplanar waveguide resonator.  \textit{Nat. Commun.} \textbf{8}, 2254 (2017).
\bibitem{FortaghPRapp2025} Wilde, B. et al. Superconducting on-chip microwave cavity for tunable hybrid systems with optically trapped Rydberg atoms.  \textit{Phys. Rev. Applied} \textbf{23}, 064016 (2025).
\bibitem{Kumar2023} Kumar, A. et al. Quantum-enabled millimetre wave to optical transduction using neutral atoms.  \textit{Nature}  \textbf{615}, 614 (2023).}





\end{thebibliography}
\end{document}